\documentclass[english,aps,prl,twocolumn,amsmath,amssymb,showpacs,superscriptaddress,notitlepage,longbibliography,nofootinbib]{revtex4-2}
\usepackage[T1]{fontenc}
\usepackage{color}
\usepackage{amsmath}
\usepackage{graphicx}
\usepackage{esint}
\usepackage{tikz}
\usepackage{bbm}
\usepackage{bm}

\usepackage{pdfpages}
\makeatletter
\AtBeginDocument{\let\LS@rot\@undefined}
\makeatother

\usepackage[utf8]{inputenc}

\newcommand{\calW}{\mathcal{W}} 

\newcommand{\calL}{\mathcal{L}}
\newcommand{\calA}{\mathcal{A}}
\newcommand{\calB}{\mathcal{B}}
\newcommand{\calG}{\mathcal{G}}
\newcommand{\calH}{\mathcal{H}}

\DeclareMathOperator\bfk{{\bf k}} 

\makeatletter
 
\usepackage{amssymb}
\usepackage{amsmath}
\usepackage{graphicx}
\usepackage[colorlinks=true,linkcolor=blue,anchorcolor=red,citecolor=blue,urlcolor=blue]{hyperref}
\usepackage[caption=false]{subfig}
\usepackage{lipsum}

\makeatother

\usepackage{babel}
\begin{document}
\renewcommand{\figurename}{Fig.}
\title{Non-Hermitian spatial symmetries and their stabilized normal and exceptional topological semimetals}

\author{W.\ B. Rui}
\email{wbrui@hku.hk}
\address{Department of Physics and HKU-UCAS Joint Institute for Theoretical
and Computational Physics at Hong Kong, The University of Hong Kong,
Pokfulam Road, Hong Kong, China}

\author{Zhen Zheng}
\address{Department of Physics and HKU-UCAS Joint Institute for Theoretical
and Computational Physics at Hong Kong, The University of Hong Kong,
Pokfulam Road, Hong Kong, China}
\author{Chenjie Wang}
\email{cjwang@hku.hk}

\address{Department of Physics and HKU-UCAS Joint Institute for Theoretical
and Computational Physics at Hong Kong, The University of Hong Kong,
Pokfulam Road, Hong Kong, China}
\author{Z.\ D. Wang}
\email{zwang@hku.hk}

\address{Department of Physics and HKU-UCAS Joint Institute for Theoretical
and Computational Physics at Hong Kong, The University of Hong Kong,
Pokfulam Road, Hong Kong, China}
\date{\today}
\begin{abstract}
We study non-Hermitian spatial symmetries---a class of symmetries that have no counterparts in Hermitian systems---and study how normal and exceptional semimetals can be stabilized by these symmetries. 
Different from internal ones, spatial symmetries act nonlocally in momentum space and enforce global constraints on both band
degeneracies and topological quantities at different locations. 
In deriving general constraints on band degeneracies and topological invariants,
we demonstrate that non-Hermitian spatial symmetries are on an equal footing with, but are essentially different from Hermitian ones.
First, we discover the nonlocal Hermitian conjugate pair
of exceptional or normal band degeneracies that are enforced by non-Hermitian spatial symmetries. Remarkably,
we find that these pairs lead to the symmetry-enforced violation of the Fermion doubling theorem in the
long-time limit. Second, with the topological constraints, we unravel that certain exceptional manifold is only compatible with and stabilized by non-Hermitian spatial symmetries but is intrinsically incompatible with Hermitian spatial symmetries. We illustrate these findings using two three-dimensional models of a non-Hermitian Weyl semimetal and an exceptional unconventional Weyl semimetal. Experimental cold-atom realizations of both models are also proposed.
\end{abstract}

\maketitle

\textit{\textcolor{blue}{Introduction.}}\textit{\textemdash{}} Symmetry serves as a guiding principle in the study of topological phases.
A hallmark is the classification of topological
phases with internal symmetries
\citep{Schnyder_PRB_2008,kitaev_periodic_2009,Chiu_RMP_2016} or  spatial symmetries (i.e., topological crystalline
phases) \citep{Liang2011crystalline,tanaka_experimental_2012,dziawa_topological_2012,Ando_Topological_2015,Fang_2015_new,Khalaf_2018_symmetry,po_symmetry-based_2017,Chiu_RMP_2016,tang_comprehensive_2019,zhang_catalogue_2019,vergniory_complete_2019,Slager_PRX_2017,Benalcazar61,rui_higher-order_2020,rui_intertwined_2022}. Recently, the study has been extended into the non-Hermitian
regime \citep{zhou_observation_2018,bandres_topological_2018,zhao_non-hermitian_2019,Leykam_PRL_2017,shen_PRL_2018,PRL_2021_yang,xiao_non-hermitian_2020,Okugawa_PRB_2021,Vecsei_PRB_2021,Shiozaki_PRB_2021,rui_classification_2019}. 
In non-Hermitian systems, besides normal semimetals with nondefective degeneracies, there are exceptional semimetals
characterized by exceptional points (EPs)~\citep{kato_perturbation,ep_intro,miri_exceptional_2019,Lee-eb-prl}, at which the Hamiltonian is defective and the energy bands are also degenerate. These degeneracies may 
collectively form normal or exceptional manifolds~\citep{footnote}, such as rings, surfaces, and complex structures like a nexus~\citep{Xu_weyl_nh,Yoshida_ring_2019,cerjan_experimental_2019,Ghorashi_dirac_2021,Liu_PRL_2021,Rui_dirac_nh,Rui_dirac_nh,Jin_PRL_2019,zhou_exceptional_2019,zhang_tidal_2021,tang_exceptional_2020,Wang_PRL_2021,Bergholtz_PRA_2018,Zhang_PRA_2020}. 

Internal symmetries, although widely studied in non-Hermitian systems~\citep{Kawabata_ep_2019,non-hermi_nodal,stalhammar_classification_2021,Delplace_PRL_2021,bernard_classification_2002,non-hermi_symmetry,non-Hermi_symmetry_classes,Periodic_table_nh,non-hermi_rmp}, seem to be playing a small role in stabilizing the global configuration of the above-mentioned band degeneracies and their formed manifolds. Thus, it is then natural to resort to spatial symmetries. Like internal symmetries that are greatly ramified by non-Hermiticity~\citep{bernard_classification_2002,non-hermi_symmetry,non-Hermi_symmetry_classes,Periodic_table_nh,Li-prb-topo,non-hermi_rmp}, spatial symmetries also come in different classes, such as the Hermitian and non-Hermitian classes --- see Eqs.~\eqref{eq:hermi_SM} and \eqref{eq:nonhermi_SM} for definitions. So far, it is unclear whether and how different classes of spatial symmetries can stabilize band degeneracies and constrain topological properties in non-Hermitian systems.

\begin{figure}[t]
	\includegraphics[width=0.9\columnwidth]{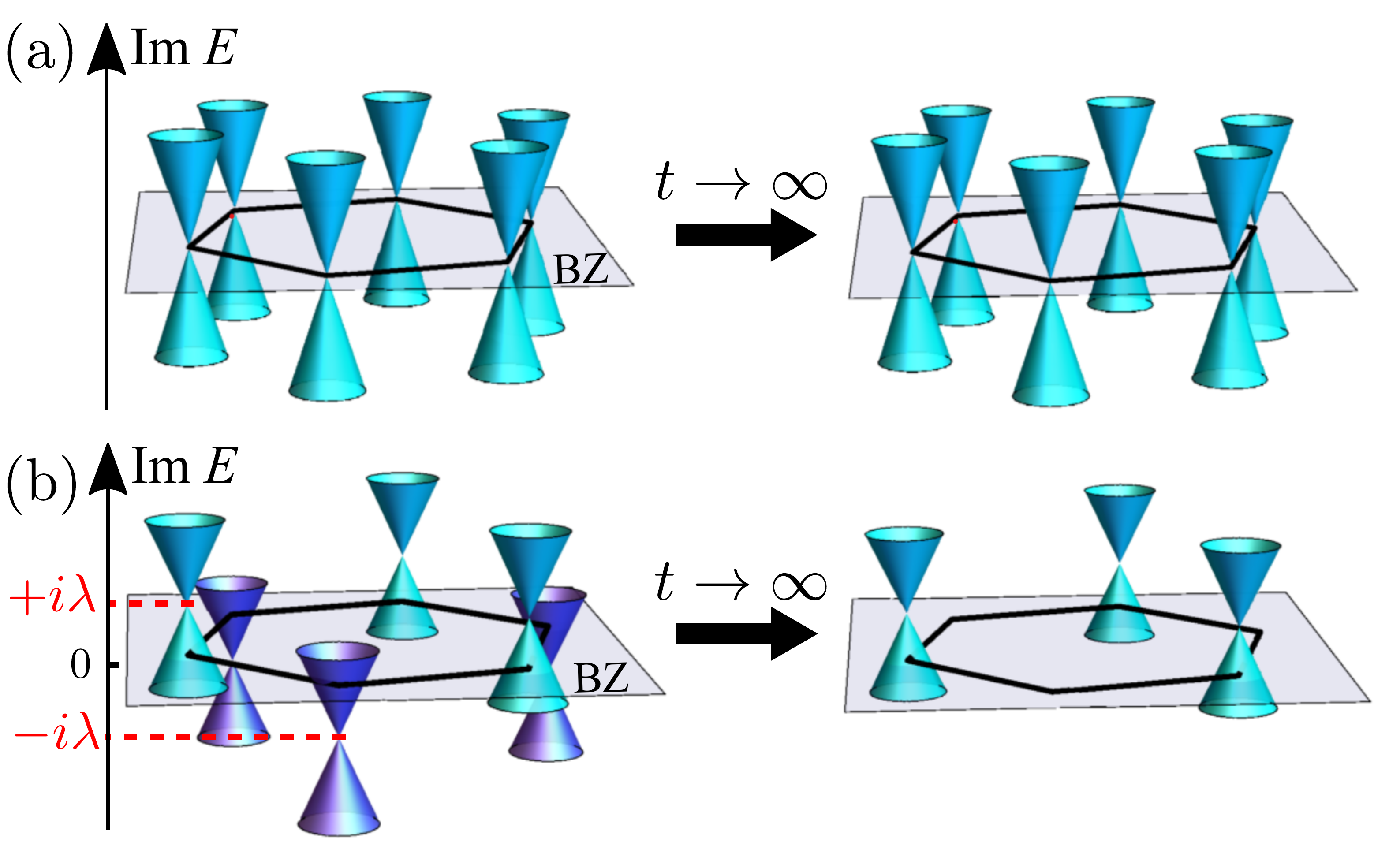}
	\caption{Schematic for the violation of the Fermion doubling theorem enforced by the non-Hermitian spatial symmetry in the long-time limit (b). For the Hermitian spatial symmetry in (a), the theorem is still respected. Here the six-fold
		rotation symmetry is taken for illustration.
		\label{fig:Fermi-doub}}
\end{figure}

In this work, we focus on non-Hermitian spatial symmetries and demonstrate how they characterize and stabilize normal and exceptional topological semimetals.
First, we show that both normal and exceptional band degeneracies are preserved under symmetry operations,
similar to Hermitian cases. But there is a stark difference: the symmetry-related band degeneracies form a nonlocal Hermitian conjugate pair in momentum space, and, thus, must possess opposite imaginary energies. As imaginary energy determines the inverse lifetime, only half of these degeneracies survive in the long-time limit, leading
to the violation of the Fermion doubling theorem, as shown in Fig.~\ref{fig:Fermi-doub}(b). Second, we show that compared to Hermitian ones, non-Hermitian spatial symmetries play an equivalent role, but act differently in constraining the  
topological quantities, including Wilson loops, Chern numbers, and winding numbers. We explore the exceptional unconventional Weyl semimetal to show that certain exceptional manifold [e.g., Fig.~\ref{fig:exp_weyl}(b)] is compatible with and can only be stabilized by non-Hermitian spatial symmetries, but it is intrinsically incompatible with Hermitian spatial symmetries due to constraints on topological quantities. We also discuss possible realizations of our models in cold-atomic systems.

\textit{\textcolor{blue}{Non-Hermitian spatial symmetries.}}\textit{\textemdash{}}
Consider a crystal or lattice system with lattice-translation symmetries,
so that the Hamiltonian can be transformed into momentum space. If the system respects a crystalline symmetry, the Hamiltonian usually transforms as
\begin{equation}
    \text{\ensuremath{\mathcal{G}}\ensuremath{\ensuremath{\mathcal{H}}}}(\mathbf{k})\mathcal{\mathcal{G}}^{-1}=\text{\ensuremath{\mathcal{H}}}(g\mathbf{k}),
    \label{eq:hermi_SM}
\end{equation}
where we take $\mathcal{G}$ to be unitary and $g$ transforms the crystal momentum $\mathbf{k}$. In Hermitian systems, the above transformation is equivalent to
\begin{equation}
\text{\ensuremath{\mathcal{G}}\ensuremath{\ensuremath{\mathcal{H}}}}(\mathbf{k})\mathcal{\mathcal{G}}^{-1}=\text{\ensuremath{\mathcal{H}}}^{\dagger}(g\mathbf{k}).\label{eq:nonhermi_SM}
\end{equation}
However, the equivalence no longer holds in non-Hermitian systems as $\text{\ensuremath{\mathcal{H}}}^{\dagger}(\mathbf{k})\neq\text{\ensuremath{\mathcal{H}}}(\mathbf{k})$. Accordingly, Eqs.~\eqref{eq:hermi_SM} and \eqref{eq:nonhermi_SM} describe different classes of symmetries.
Such a ramification by non-Hermiticity is similar to that of
nonspatial symmetries. The latter has been systematically studied, e.g., in Ref.~\citep{non-hermi_symmetry}, which shows that there are 38-fold symmetry classes, 
far beyond the celebrated Altland-Zirnbauer 10-fold classes in Hermitian systems. We will refer to those satisfying \eqref{eq:hermi_SM} as ``Hermitian spatial symmetries'', and those satisfying \eqref{eq:nonhermi_SM} as ``non-Hermitian spatial symmetries''. Note that $\mathcal{G}$ can also be anti-unitary. However, we focus on the unitary case below.

Let $|\Psi_{L,n}(\mathbf{k})\rangle$ and $|\Psi_{R,n}(\mathbf{k})\rangle$ be the left and right eigenvectors of the non-Hermitian Hamiltonian $\ensuremath{\mathcal{H}}(\mathbf{k})$, respectively, where $n$ is the band index.
The two vectors satisfy $\text{\ensuremath{\mathcal{H}}(\ensuremath{\mathbf{k}})}|\Psi_{R,n}(\mathbf{k})\rangle =E_{n}(\mathbf{k})|\Psi_{R,n}(\mathbf{k})\rangle$ and $
\text{\ensuremath{\mathcal{H}}}^{\dagger}(\mathbf{k})|\Psi_{L,n}(\mathbf{k})\rangle =E_{n}^{*}(\mathbf{k})|\Psi_{L,n}(\mathbf{k})\rangle$. They are different in general and form a
biorthonormal basis, satisfying $\langle \Psi_{L,m}(\mathbf{k})|\Psi_{R,n}(\mathbf{k})\rangle=\delta_{mn}$. If $\mathcal{H}(\mathbf{k})$ admits a non-Hermitian spatial symmetry $\mathcal{G}$ satisfying Eq.~(\ref{eq:nonhermi_SM}), one can show that
\begin{align}
    &\text{\ensuremath{\mathcal{H}}}^{\dagger}(g\mathbf{k})\mathcal{G}|\Psi_{R,n}(\mathbf{k})\rangle=E_{n}(\mathbf{k})\mathcal{G}|\Psi_{R,n}(\mathbf{k})\rangle,\nonumber \\
    &\text{\ensuremath{\mathcal{H}}}(g\mathbf{k})\mathcal{G}|\Psi_{L,n}(\mathbf{k})\rangle=E_{n}^*(\mathbf{k})\mathcal{G}|\Psi_{L,n}(\mathbf{k})\rangle.
\end{align}
Accordingly, every right (left) eigensystem $\{|\Psi_{R,n}(\mathbf{k})\rangle,E_n(\mathbf{k})\}$ $\left(\{|\Psi_{L,n}(\mathbf{k})\rangle,E_n^*(\mathbf{k})\}\right)$ at $\mathbf{k}$ is mapped to a left (right) eigensystem $\{\mathcal{G}|\Psi_{R,n}(\mathbf{k})\rangle,E_n^*(\mathbf{k})\}$ $\left(\{\mathcal{G}|\Psi_{L,n}(\mathbf{k})\rangle,E_n(\mathbf{k})\}\right)$ at  $g\mathbf{k}$. As $\mathcal{G}$ is invertible, this map is a one-to-one correspondence. On the other hand, a Hermitian symmetry $\calG$ maps right (left) eigensystem to right (left) eigensystem. 

\textit{\textcolor{blue}{Nonlocal Hermitian conjugate pair of normal or exceptional degeneracies.}}\textit{\textemdash{}} The non-Hermitian symmetry $\mathcal{G}$ also
maps between band degeneracies nonlocally, but in a way different from Hermitian one. Consider a
degeneracy at momentum $\mathbf{k_{\text{D}}}$. The Hamiltonian
$\mathcal{H}(\mathbf{k}_D)$ can be transformed by an invertible
matrix $\mathcal{P}$ as
\begin{align}\label{eq:deg}
\mathcal{P}^{-1}\mathcal{H}(\mathbf{k}_D)\mathcal{P}=\mathcal{J}(\mathbf{k}_D),\quad \mathcal{J}(\mathbf{k}_D)=E(\mathbf{k}_D)\mathbbm{1}+\sigma\mathcal{N}.
\end{align}
Here, $\mathbbm{1}$ is the identity matrix and $\mathcal{N}$ the nilpotent matrix defined by $\mathcal{N}_{ij}=\delta_{i,j-1}$. $\sigma=0$ corresponds to normal degeneracies, and $\sigma=1$ corresponds to a Jordan block for exceptional degeneracies, i.e., EPs. 
In the presence of a non-Hermitian symmetry (\ref{eq:nonhermi_SM}), it can be derived that
\begin{equation}
\tilde{\mathcal{P}}\mathcal{H}(g\mathbf{k}_D)\tilde{\mathcal{P}}^{-1}=\mathcal{J}^{\dag}(\mathbf{k}_D),\label{eq:gEP}
\end{equation}
where $\tilde{\mathcal{P}}=\mathcal{P^{\dagger}\mathcal{G}^{\dagger}}$
is invertible. By comparing Eqs.~\eqref{eq:deg} and \eqref{eq:gEP}, 
we can see that $\mathcal{H}(g\mathbf{k}_D)$
is brought to the same $\mathcal{J}(\mathbf{k}_D)$ as $\mathcal{H}(\mathbf{k}_D)$, meaning that the normal or exceptional degeneracy is preserved under the symmetry operation. However, a stark difference is the Hermitian conjugation
in \eqref{eq:gEP}, which does not appear for the Hermitian spatial symmetry.
It makes the symmetry-related degeneracies form nonlocal Hermitian 
conjugate pairs in momentum space.

\textit{\textcolor{blue}{Symmetry-enforced violation of Fermion doubling theorem in the long-time limit.}}\textit{\textemdash{}} For topological point degeneracies, such as Weyl points, they must come in pairs on lattice according to
the Fermion doubling theorem~\citep{nielsen_no-go_1981,nielsen_absence_1981,nielsen_absence_1981-1}. Recently, this concept has been extended to include EPs in non-Hermitian systems~\citep{Chiu-NH_Double}. Under non-Hermitian spatial symmetry, as the two symmetry-related normal or exceptional band degeneracies form a nonlocal Hermitian conjugate pair, their energies obey
\begin{equation}\label{energy-relation}
E(g\mathbf{k}_{D}) = E^*(\mathbf{k}_{D}).  
\end{equation}
It means that the two degeneracies are distinguishable by
their opposite imaginary energies, as shown by the blue and cyan cones in the left panel of Fig.~\ref{fig:Fermi-doub}(b). Such a distinction
is not possible under Hermitian spatial symmetry with $E(g\mathbf{k}_{D}) = E(\mathbf{k}_{D})$, as shown in (a). 

Remarkably, the above symmetry-enforced separation of imaginary energy between degeneracies leads to anomalous behaviors. In the long-time limit~\citep{non-hermi_rmp,Lee-dynamics}, as the imaginary
energy determines the inverse lifetime, only the modes at
the degeneracies with positive imaginary energies survive. Thus, effectively, only half of the symmetry-related degeneracies (normal or exceptional) exist, leading to the violation of Fermion doubling theorem in the long-time limit, as shown in Fig.~\ref{fig:Fermi-doub}(b). 
Note that this cannot happen for a Hermitian 
symmetry $\mathcal{G}$.

\textit{\textcolor{blue}{Topological quantities: Wilson loop, Chern number, and winding number.}}\textit{\textemdash{}}
The non-Hermitian spatial symmetries play an equivalent role but act in a different way than Hermitian spatial symmetries in constraining topological
quantities~\citep{dai2011prb,Bernevig2014prb,Soluyanov2016prx,hughes2017science}.
For biorthonormal eigenstates, Wilson loops in
non-Hermitian systems can be defined as~\citep{luo2019prl,Zhao2021prl}:
\begin{equation}\label{wilson-def}
\mathcal{W}^{\alpha\bar{\alpha}}_{\mathcal{L}}=\overline{\exp} \left[-\oint_{\mathbf{k}_0} d\mathbf{k} \mathcal{A}^{\alpha\bar{\alpha}}(\mathbf{k})\right],    
\end{equation} 
where $\alpha= R,L$ (with $\bar{R}=L$ and $\bar{L}=R$), $\calL$ is a loop in momentum space with $\bfk_0$ being a base point, and  "$\overline{\exp}$" denotes that the integral is path ordered. The non-Abelian Berry connection is defined as
$\mathcal{A}^{\alpha\bar{\alpha}}_{mn}(\mathbf{k})=\langle \Psi_{\alpha,m}(\mathbf{k})|\partial_\mathbf{k}|\Psi_{\bar{\alpha},n}(\mathbf{k})\rangle$ for a set of bands that are separated from other bands along the loop $\calL$. The Wilson loop $\calW^{\alpha\bar{\alpha}}_\calL$ is invariant under a basis transformation (gauge transformation) only in the Abelian case (i.e., a single band). For multiple bands, one needs to consider the determinant
\begin{align}\label{wilson-phase}
    \det\left(\calW^{\alpha\bar{\alpha}}_\calL\right) = \exp\left(a_\calL^{\alpha\bar{\alpha}} + i \gamma_\calL^{\alpha\bar{\alpha}}\right),
\end{align}
where both $a_\calL^{\alpha\bar{\alpha}} $ and $ \gamma_\calL^{\alpha\bar{\alpha}}$ are real. The phase $\gamma_\calL^{\alpha\bar{\alpha}}$ is the Berry phase. We show in the Supplemental Material (SM) \citep{SuppInf} that 
$a_\calL^{LR} = -a_\calL^{RL}$,   $\gamma_\calL^{LR}=\gamma_\calL^{RL}$.

With a non-Hermitian spatial symmetry in \eqref{eq:nonhermi_SM}, the Wilson loop satisfies the following relation (see SM \citep{SuppInf} for details)
\begin{equation}\label{wilson-relation}
\mathcal{W}^{\alpha\bar{\alpha}}_{\mathcal{L}} =\mathcal{S}_{g,\alpha}^\dagger(\mathbf{k}_0) \tilde{\mathcal{W}}^{\bar{\alpha}\alpha}_{g\mathcal{L}}\mathcal{S}_{g,\bar{\alpha}}(\mathbf{k}_0),
\end{equation}
where $g\calL$ is the image of $\calL$ under $\calG$, and the sewing matrix 
$\mathcal{S}_{g,\alpha}^{\tilde{n}n}(\mathbf{k})=\langle \Psi_{\alpha,
\tilde{n}}(g\mathbf{k})| \mathcal{G}|\Psi_{\alpha,n}(\mathbf{k})\rangle$. Here, ``$\tilde{n}$'' indexes the bands associated with the states $\mathcal{G}|\Psi_{\alpha,n}(\mathbf{k})\rangle$, which are not necessarily the same as those of $|\Psi_{\alpha,n}(\mathbf{k})\rangle$, and $\tilde{\calW}_{g\calL}^{\alpha\bar{\alpha}}$ is the corresponding Wilson loop. Unitarity of $\calG$ leads to
$\mathcal{S}_{g,\alpha}^\dagger(\mathbf{k})\mathcal{S}_{g,\bar{\alpha}}(\mathbf{k})=\mathbbm{1}$. Taking the determinant on both sides of \eqref{wilson-relation}, we obtain
$\tilde{a}_{g\calL}^{\alpha\bar{\alpha}}= a_\calL^{\bar{\alpha}\alpha}$,  $\tilde{\gamma}_{g\calL}^{\alpha\bar{\alpha}} = \gamma_\calL^{\bar{\alpha}\alpha}$.
Instead, if $\calG$ is a Hermitian symmetry, we have   $\tilde{a}_{g\calL}^{\alpha\bar{\alpha}}= a_\calL^{\alpha\bar{\alpha}}$ and $\tilde{\gamma}_{g\calL}^{\alpha\bar{\alpha}}= \gamma_\calL^{\alpha\bar{\alpha}}$.

Chern numbers can also be defined in non-Hermitian systems. The non-Abelian Berry curvature is defined as $\calB^{\alpha\bar{\alpha}} = i\nabla \times \calA^{\alpha\bar{\alpha}} + i \calA^{\alpha\bar{\alpha}}\times\calA^{\alpha\bar{\alpha}}$. Then, associated with every closed surface $\Sigma$ on which a set of energy bands are separate from others, the Chern number is given by
\begin{align}\label{chern-def}
    C_\Sigma = \frac{1}{2\pi} \mathrm{Re} \int_\Sigma d\mathbf{S}\cdot \mathrm{tr}\left(\calB^{\alpha\bar{\alpha}}\right).
\end{align}
We show in the SM \citep{SuppInf} that $C_\Sigma$ is independent of $\alpha$ and that it takes integer values. Similar to the Berry phase, one can show that $C_\Sigma =\tilde C_{g\Sigma}$, where $g\Sigma$ is the image of $\Sigma$ under $\calG$, and $\tilde{C}_{g\Sigma}$ is associated with the bands of $\calG|\Psi_{R,n}(\bfk)\rangle$.

\begin{table}[t]
	\centering 
	\caption{Transformation rules of different topological quantities under Hermitian and non-Hermitian spatial symmetries. We have denoted $\tilde{W}\equiv W(g\bfk_{\rm EP})$ and $W\equiv W(\bfk_{\rm EP})$, see Eq.~\eqref{eq:inv_connected}. Here, no. is short for number.}
	\begin{tabular}{c c c}
		\hline\hline
		& Hermitian & Non-Hermitian \\
		\hline
		Wilson loop &  $\tilde{a}_{g\calL}^{\alpha\bar{\alpha}}={a}_{\calL}^{\alpha\bar{\alpha}}$ & $\tilde{a}_{g\calL}^{\alpha\bar{\alpha}}={a}_{\calL}^{\bar{\alpha}\alpha}$ \\
		&  $\tilde{\gamma}_{g\calL}^{\alpha\bar{\alpha}}=\gamma_{\calL}^{\alpha\bar{\alpha}}$ & $\tilde{\gamma}_{g\calL}^{\alpha\bar{\alpha}}=\gamma_{\calL}^{\bar{\alpha}\alpha}$ \\
		Chern no. & $\tilde C_{g\Sigma}=C_\Sigma$ & $\tilde C_{g\Sigma}=C_\Sigma$ \\
		Winding no. & $\tilde W=+\sigma(\mathbf{k}_\text{EP})\sigma(g) W$ & $\tilde W=-\sigma(\mathbf{k}_\text{EP})\sigma(g) W$ \\
		\hline\hline
	\end{tabular}
	\label{Table:topo-quant}
\end{table}

We now turn to the topology of EPs. As proved in the SM~\citep{SuppInf}, for order-2 EPs in three-dimensional (3D) systems, they generally form exceptional lines (ELs).
Assuming a gap around such an EL, a winding number can be defined~\citep{Kawabata_ep_2019,shen_PRL_2018}:
\begin{equation}
W(\mathbf{k}_{\text{EP}})=\frac{1}{2\pi i}\oint_{S^{1}}d\mathbf{k}\cdot\nabla_{\mathbf{k}}\log\det\left[\mathcal{H}(\mathbf{k})-E(\mathbf{k}_{\text{EP}})\right],\label{eq:top_inv}
\end{equation}
where $S^{1}$ is a loop that encircles the EL and $\mathbf{k}_{\rm EP}$ is any point on the EL. The integral \eqref{eq:top_inv} needs an orientation on $S^1$ to be unambiguous. It can be done by first assigning an orientation to the EL, which then induces a orientation on $S^1$ through the right-hand rule. Changing the orientation of EL gives a minus sign to $W(\mathbf{k}_{\text{EP}})$. 

In the presence of spatial symmetry $\calG$, 
we show in the SM ~\citep{SuppInf} that
\begin{align}
W(g\mathbf{\mathbf{k}_{\text{EP}}})=\zeta\sigma( \mathbf{k}_{\rm EP})\sigma(g) W(\mathbf{\mathbf{k}_{\text{EP}}}),\label{eq:inv_connected}
\end{align}
where $\zeta=+1$ if $\calG$ is a Hermitian symmetry, and $\zeta=-1$ if $\calG$ is a non-Hermitian symmetry. The factor $\sigma( \mathbf{k}_{\rm EP})=\sigma(g\mathbf{\mathbf{k}_{\rm EP}}, \mathbf{k}_{\rm EP})= 1$ if the orientations of the ELs at $\mathbf{k}_{\text{EP}}$ and $g\mathbf{k}_{\text{EP}}$ match under $\calG$, and $\sigma(g\mathbf{\mathbf{k}_{\rm EP}}, \mathbf{k}_{\rm EP})= -1$ otherwise. The factor $\sigma(g)=1$ or $-1$, if $\calG$ preserves (e.g., rotation) or reverses (e.g., mirror reflection) the chirality of the momentum space, respectively.

\begin{figure}[b]
	\includegraphics[width=0.9\columnwidth]{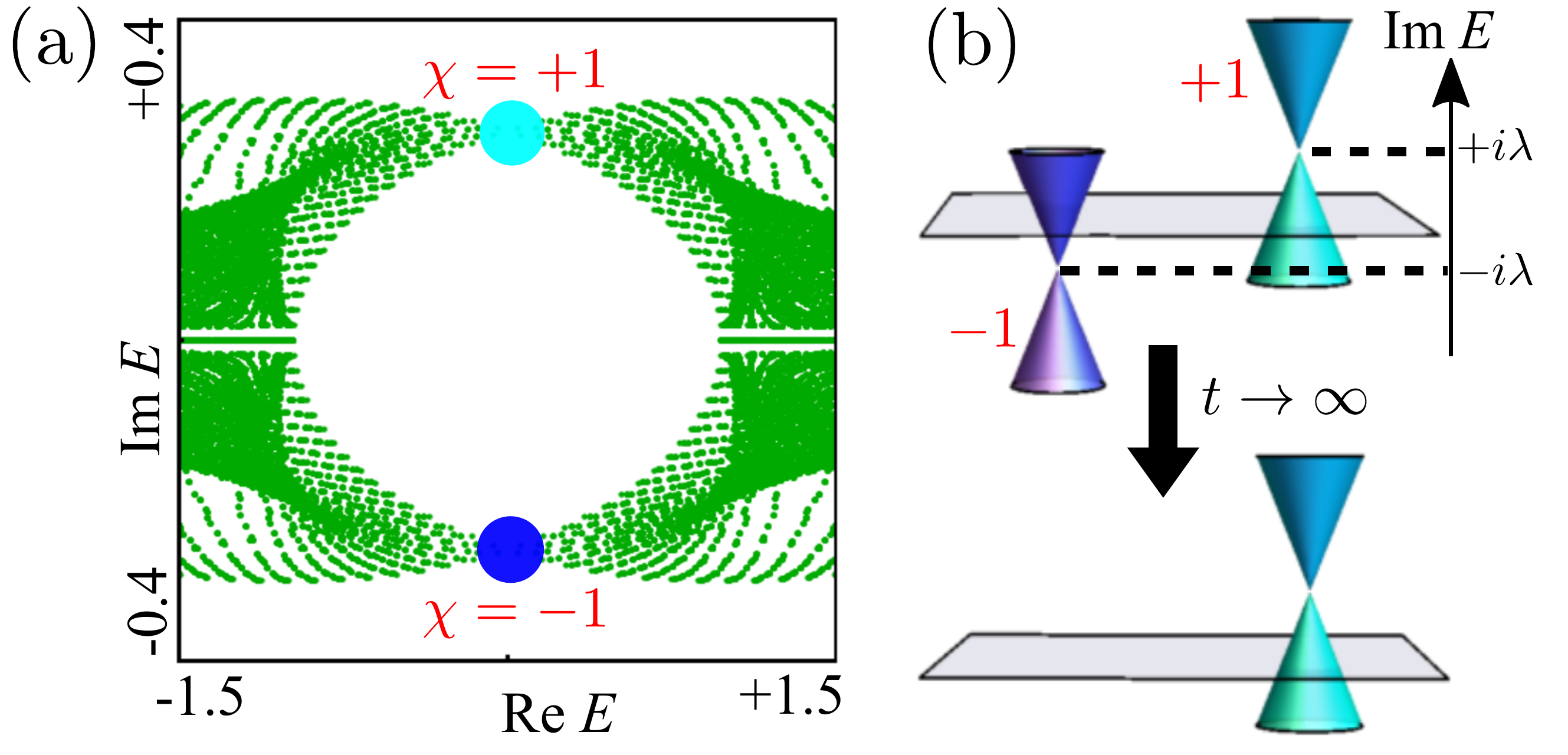}
	\caption{(a) The energy spectrum for the Weyl semimetal protected by the non-Hermitian inversion symmetry. (b) The two Weyl points possess opposite imaginary energies due to the symmetry, which leads to a single Weyl point on lattice in the long-time limit. The parameters in the model~\eqref{weyl-Fermion} are $A=t=k_0=1.0, M=3.0$, and $\lambda=0.3$.
	\label{fig:WP}}
\end{figure}

\textit{\textcolor{blue}{Non-Hermitian Weyl semimetals.}}\textit{\textemdash{}} With the above general results, we proceed to discuss two concrete models.  The first model is a normal Weyl 
semimetal protected by a non-Hermitian inversion symmetry. In the long-time limit, there is only one Weyl point in effect, violating the Fermion doubling theorem. 
The cold-atom realization of this model can be found in the SM~\citep{SuppInf}. The model Hamiltonian in momentum space reads
\begin{align}\label{weyl-Fermion}
\mathcal{H}(\mathbf{k})=&A\sin k_x \tau_1\sigma_1+A\sin k_y \tau_1\sigma_2+A\sin k_z\tau_1 \sigma_3 \nonumber \\
+&M(\mathbf{k})\tau_3\sigma_0+k_0\tau_0\sigma_3+i\lambda\tau_1\sigma_0,
\end{align}
where $M(\mathbf{k})=(t\cos k_x+t\cos k_y+t\cos k_z-M)$ and $M,A,t, k_0, \lambda$ are real  parameters. The system respects the non-Hermitian inversion symmetry $\mathcal{I}=\tau_3\sigma_0$ as
\begin{align}\label{nh-inv}
\mathcal{I}\mathcal{H}(\mathbf{k})\mathcal{I}^{-1}=\mathcal{H}(-\mathbf{k})^\dagger.
\end{align}
As shown in Fig.~\ref{fig:WP}, for the chosen parameters, this model features two Weyl points whose low-energy effective models can be obtained
as
\begin{align}
    &h_\text{WP1}(\delta\mathbf{k})=+\delta\mathbf{k}\cdot\bm{\sigma}+i\lambda\sigma_0,\quad \mathbf{k}=(0,0,-k_0);\nonumber\\
    &h_\text{WP2}(\delta\mathbf{k})=-\delta\mathbf{k}\cdot\bm{\sigma}-i\lambda\sigma_0,\quad \mathbf{k}=(0,0,+k_0).
\end{align}
As the two Weyl points with opposite chiralities ($\chi=\pm1$) are connected by the non-Hermitian symmetry, they form a nonlocal Hermitian conjugate pair of normal degeneracies and possess opposite imaginary energies of $\pm i\lambda$.

We use the time-evolution operator $\mathcal{U}(\mathbf{k},t)= \mathcal{T}\exp{[-i/\hbar \int_{0}^{t}}\mathcal{H}(\mathbf{k})dt']$ ($\mathcal{T}$: time ordering), which is generally not unitary in non-Hermitian systems~\citep{Ashida_ap_2020}, to investigate the dynamics. After a 
sufficiently long time, i.e., $t\gg \hbar/\lambda$, the Weyl point described by $h_\text{WP2}(\delta\mathbf{k})$ with $-i\lambda$ vanishes due to the exponentially decaying factor in its time-evolution operator. Thus,
in the long-time limit, only the Weyl point described by $h_\text{WP1}(\delta\mathbf{k})$ with imaginary energy $+i\lambda$ survives, leading to the violation of Fermion doubling theorem, as shown in Fig.~\ref{fig:WP}(b).

\textit{\textcolor{blue}{Exceptional unconventional Weyl semimetals.}}\textit{\textemdash{}} Next, we study a non-Hermitian extension of unconventional Weyl semimetals, where the momentum space hosts monopoles of charge $\pm2$. This model demonstrates how non-Hermitian spatial symmetries can stabilize exceptional manifolds in a different way from the Hermitian ones, and illustrate the transformation rules of topological quantities in Table~\ref{Table:topo-quant}.  We discuss a cold-atom realization of this model in the SM~\citep{SuppInf}.

\begin{figure}[t]
\includegraphics[width=0.9\columnwidth]{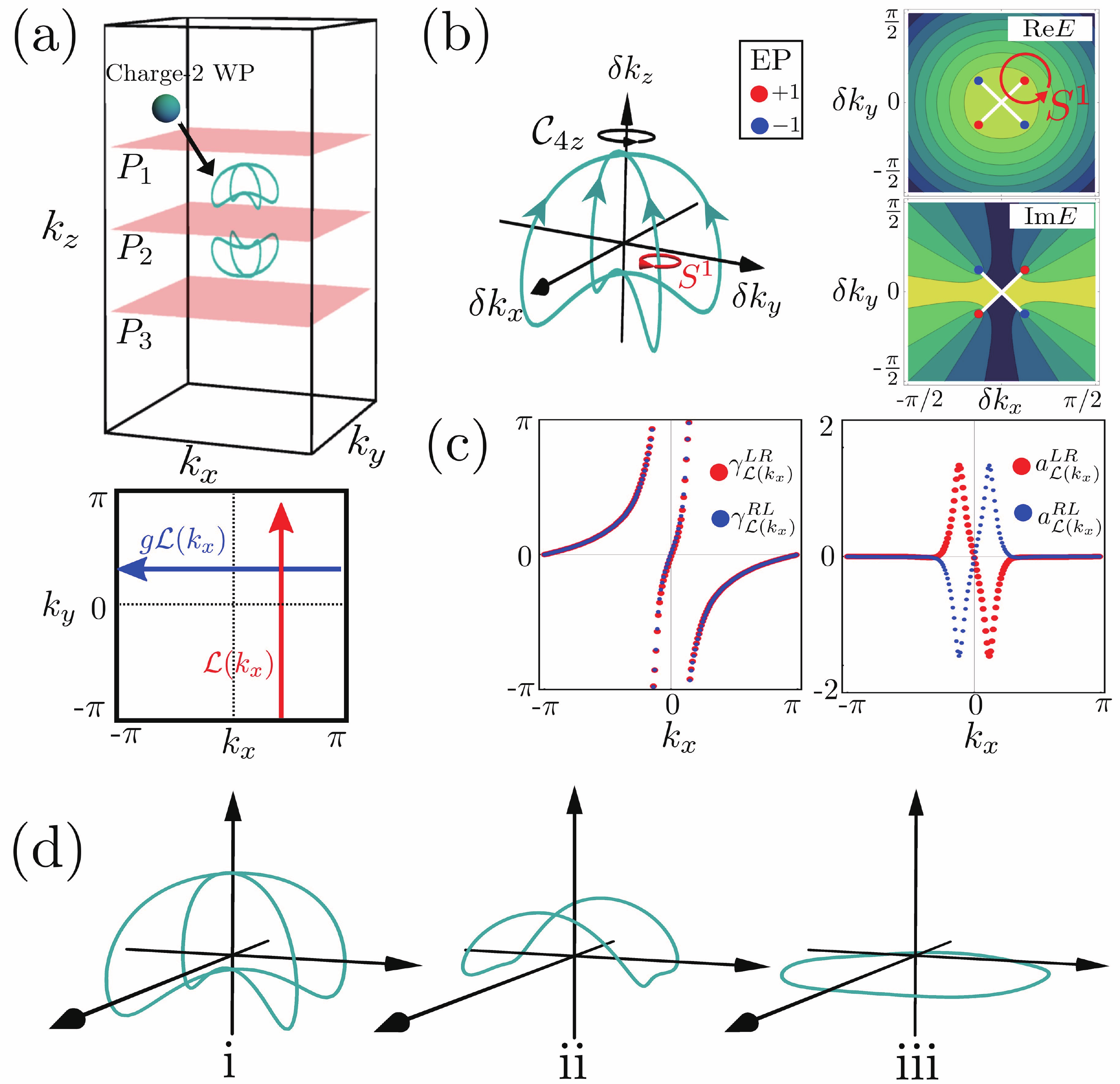}
\caption{(a) The non-Hermitian spatial symmetry preserving term turns an
unconventional Weyl point into exceptional lines (cyan lines). The lower panel shows the path (red and blue arrows) for calculating the Wilson loop for $\mathcal{L}(k_x)$ and symmetry-related $g\mathcal{L}(k_x)$ on the $P_2$ plane.
(b) Left: Zoom-in of the exceptional manifold (cyan) around $\mathbf{K}_+$. Right:  Contour plots of the real (upper) and imaginary (lower) eigenenergies on the $k_{x}k_{y}$-plane at $\delta k_{z}=0$. Here blue and red points are EPs with $W(\mathbf{k_{EP}})=+1$ and $-1$, respectively. (c) The Berry phase $\gamma_\calL^{\alpha\bar{\alpha}}$ and the real exponent $a_\calL^{\alpha\bar{\alpha}}$ for a family of non-contractible loops $\mathcal{L}: (k_x,-\pi,0)\rightarrow (k_x,\pi,0)$ with $k_x\in[-\pi, \pi]$ on the $P_2$ plane in (a), i.e., the red arrow.  (d) Evolution of the exceptional manifold (cyan) for $\calH^{w'}(\bfk)$ in \eqref{weyl2} with $\eta\in [0,1]$. The rotation $\mathcal{C}_{4z}$ is a non-Hermitian  symmetry at $\eta=0$ (i), a Hermitian  symmetry at $\eta=1$ (iii), and broken when $0<\eta<1$ (ii). The parameters in the model~\eqref{eq:exp_weyl} are $A=t_z=t_\parallel=1.0$, $M_{0}=5.5$, and $\lambda=0.3$.
\label{fig:exp_weyl}}
\end{figure}

The momentum-space Hamiltonian of our model reads
\begin{align}
\mathcal{H}^w(\mathbf{k})  = & \mathcal{H}^w_0(\mathbf{k}) +\calH_1^w(\bfk), \nonumber\\
\mathcal{H}^w_0(\mathbf{k})  = & 2A(\cos k_{x}-\cos k_{y})\sigma_{1}+2A\sin k_{x}\sin k_{y}\sigma_{2} \nonumber\\
 & +[M_{0}-2t_\parallel (\cos k_x+\cos k_y)-2t_z\cos k_{z}]\sigma_{3}, \nonumber \\
\calH_1^w(\bfk) =  & i\lambda\sigma_1 \label{eq:exp_weyl},
\end{align}
where all parameters $M_0, t_\parallel,t_z, A$ and $\lambda$ are real. The unperturbed Hamiltonian $\mathcal{H}^w_0$ respects a normal
four-fold rotation symmetry $\mathcal{C}_{4z}=\sigma_{3}$,  $\mathcal{C}_{4z}\mathcal{H}_0^w(\mathbf{k})\mathcal{C}_{4z}^{-1}=\mathcal{H}_0^w(R_{4z}\mathbf{k})$
with $R_{4z}(k_{x},k_{y},k_{z})=(k_{y},-k_{x},k_{z})$. It has two Weyl points with monopole charge $\pm2$, located at $\mathbf{K_{\pm}}=[0,0,\pm\arccos (M_{0}-4t_\parallel)/2t_z]$, which are stabilized by the Hermitian $\mathcal{C}_{4z}$ symmetry \citep{Fang2012PRL}.
With the non-Hermitian $\calH_1^w$ included, the Hermitian symmetry is broken as $[\mathcal{C}_{4z},i\lambda\sigma_{1}]\neq0$. However, as a non-Hermitain symmetry, $\mathcal{C}_{4z}$ is still
preserved, which reads
\begin{equation}
\mathcal{C}_{4z}\mathcal{H}^w(k_{x},k_{y},k_{z})\mathcal{C}_{4z}^{-1}=\mathcal{H}^w(k_{y},-k_{x},k_{z})^{\dagger}.\label{eq:exp_weyl_sym}
\end{equation}

Figure \ref{fig:exp_weyl}(a) shows the exceptional manifold of the model, and (b) shows an enlargement around $\mathbf{K_{+}}$. Each of the original Weyl points turns into four rotation-symmetric ELs that jointly terminate on the rotation axis. All EPs including those on the axis are of order 2, with winding number of $W(\mathbf{k}_{\text{EP}})=\pm 1$. As shown by the right panel of (b), the energy spectrum and EPs in the 2D $k_{x}k_{y}$
Brillouin zone (BZ) across $\mathbf{K}_+$
clearly exhibit a four-fold rotation symmetry. 
The relations below \eqref{wilson-phase} are explicitly verified for a family of loops in (c). The transformation rules of Wilson loops under non-Hermitian spatial symmetry in Table~\ref{Table:topo-quant} are also verified for the two loops related by $\mathcal{C}_{4z}$ symmetry [red and blue arrows in Fig.~\ref{fig:exp_weyl}(a) lower panel], which can be found in the SM~\citep{SuppInf}.
We further compute the Chern numbers on planes $\Sigma=P_1$, $P_2$, and $P_3$, shown by pink planes in Fig.~\ref{fig:exp_weyl}(a), which
are $C_{P_1}=0$, $C_{P_2}=-2$, and $C_{P_3}=0$, respectively. It implies that, between $P_1$ and $P_2$ and between $P_2$ and $P_3$, there must exist regions that are energetically degenerate, whose stability is guaranteed by the Chern number.

In fact, stability of the structure of this exceptional manifold---specifically, existence of the intersection points on the rotation axis---can be argued further at a topological level. Let us consider a modified Hamiltonian
\begin{align}
\mathcal{H}^{w'}(\mathbf{k})  = & \mathcal{H}^w_0(\mathbf{k}) +i\lambda [(1-\eta)\sigma_1+\eta\sigma_3].\label{weyl2}
\end{align}
When $\eta=0$, $\mathcal{H}^{w'}$ reduces to $\calH^w$. When $\eta=1$, the non-Hermitian perturbation $i\lambda\sigma_3$ respects $\mathcal{C}_{4z}$ as a Hermitian symmetry. Accordingly, by tuning $\eta\in[0,1]$, we achieve a transition from a non-Hermitian $\mathcal{C}_{4z}$ to a Hermitian $\mathcal{C}_{4z}$. When $0<\eta<1$, $\mathcal{C}_{4z}$ is not respected either as Hermitian or non-Hermitian symmetry.  

Figure~\ref{fig:exp_weyl}(d) shows the evolution of the exceptional manifold as $\eta$ varies. It exhibits a fourfold rotation symmetry both at $\eta=0$ and $\eta=1$. The key difference is that:  there are exceptional intersection points on the rotation axis at $\eta=0$, while the whole axis is non-degenerate at $\eta=1$. This difference can be explained by the constraint \eqref{eq:inv_connected}. In order for ELs to terminate on the rotation axis, the total winding number must vanish such that
\begin{align}
  \sum_{n=0}^3 W(g^n\bfk_{\rm EP})  = 0.
  \label{eq:W-rot}
\end{align}
If $\mathcal{C}_{4z}$ is Hermitian,  Eqs.~\eqref{eq:inv_connected} and \eqref{eq:W-rot} together lead to $W(\bfk_{\rm EP})=0$. In other words, the exceptional manifold in Fig.~\ref{fig:exp_weyl}[d(i)] is intrinsically incompatible with Hermitian symmetries. On the other hand, Eq.~\eqref{eq:W-rot} is always satisfied for a non-Hermitian $\mathcal{C}_{4z}$ due to Eq.~\eqref{eq:inv_connected}. Moreover, the exceptional intersection points are indeed protected by the non-Hermitian $\mathcal{C}_{4z}$~\citep{SuppInf}. Thus, we conclude that non-Hermitian and Hermitian spatial symmetries may stabilize exceptional manifolds in very different manners.

\textit{\textcolor{blue}{Discussions.}}\textit{\textemdash{}} Topological semimetals stabilized by non-Hermitian spatial symmetries provide a novel platform for investigating anomalous behaviors of unpaired normal or exceptional point degeneracies. An unpaired point degeneracy, such as a single Dirac point on the 3D topological insulator surface, represents an anomaly that may have unusual physical consequences~\citep{Fu2007PRL,Qi2008PRB,Ryu2012PRB}. We have demonstrated a mechanism for dynamically achieving unpaired point degeneracies, distinct from previous approaches using topological surface states~\citep{Fu2007PRL,denner2021natcomm}. This mechanism could be readily employed, e.g., in photonic experiments where EPs have been realized~\citep{zhou_observation_2018}.

Finally, we remark that the two Weyl models are realizable not only in cold atoms, but also in other platforms like electrical circuits and photonics. We choose cold atoms for two reasons. First, cold atoms have the advantage for investigating both noninteracting and interacting systems, and, thus, can extend our theory to many-particle physics~\citep{Xu2019-weyl-in-atom,Zhang_cold_atom}. The influence of the atomic many-body interaction can be manipulated further by Feshbach resonances~\citep{Kohler2006-feshbach,Chin2010-feshbach}. Second, in cold atoms, there exist mature techniques for studying time-evolution dynamics~\citep{Zurek2005-evolution,Hofferberth2007-evolution,Polkovnikov2011-evolution,Ren2022-evolution}, which is promising for realizing our theory about dynamically achieving unpaired degeneracies.

\vspace{0.3cm}

\begin{acknowledgements} W. B. R. is grateful to Moritz M. Hirschmann for 
valuable discussions. This work was supported by the Key-Area
Research and Development Program of GuangDong Province (Grant No.
2019B030330001), the CRF (Grants No. C6005-17G and No. C6009-20G) and GRF (Grant No. 17300220) of
Hong Kong, and the NSFC/RGC JRS Grant No. N\_HKU774/21. The authors also thank support from Guangdong-Hong Kong
Joint Laboratory of Quantum Matter. 
\end{acknowledgements}

\bibliographystyle{apsrev4-2}
\bibliography{Reference}

\begin{thebibliography}{88}
\bibitem [{foo()}]{footnote}%
  \BibitemOpen
  \href@noop {} {}\bibinfo {note} {Here, we abuse the term ``manifold''. A
  normal or exceptional manifold may not be a manifold in the mathematical
  sense.}\BibitemShut {Stop}%
\end{thebibliography}

\widetext
\clearpage


\section*{Supplementary material}

# Supplemental Material for
# "Non-Hermitian spatial symmetries and their stabilized normal and exceptional topological semimetals"

W. B. Rui, Zhen Zheng, Chenjie Wang, and Z. D. Wang

In this Supplemental Material, we derive Wilson loops in non-Hermitian systems in Sec. I, reveal the relations and constraints on Wilson loop and Chern number under non-Hermitian symmetry in Sec. II, calculate the relation of winding numbers between symmetry-related exceptional lines in Sec. III, discuss the symmetries in exceptional unconventional Weyl semimetals in Sec. IV, discuss another example of exceptional triple-point semimetals in Sec. V, and propose the experimental realizations of Weyl semimetals protected by non-Hermitian inversion symmetry in Sec. VI and exceptional unconventional Weyl semimetals in Sec. VII.

## I. WILSON LOOPS IN NON-HERMITIAN SYSTEMS

In this section, we discuss some general properties of Wilson loop observables in non-Hermitian band theory. While only Abelian Wilson loops are used in the main text, here we consider general non-Abelian Wilson loops associated with $N$ bands. Recall that in Hermitian systems a Wilson line is defined as

$$\mathcal{W}_{\mathcal{P}} = \overline{\exp}\left[-\int_{\mathbf{k}_i}^{\mathbf{k}_f} d\mathbf{k}\mathcal{A}(\mathbf{k})\right], \tag{1}$$

where $\mathcal{P}$ is a path from $\mathbf{k}_i$ to $\mathbf{k}_f$ in the momentum space, $\mathcal{A}(\mathbf{k})$ is the matrix-valued Berry connection, "—" indicates that the integral is path-ordered, and the $N$ bands are separated from other bands along the path $\mathcal{P}$. We will denote the path as $\mathcal{P}: \mathbf{k}_f \leftarrow \mathbf{k}_i$. Given a set of orthonormal energy eigenstates $\langle \Psi_m(\mathbf{k})|\Psi_n(\mathbf{k})\rangle = \delta_{mn}$, the Berry connection is defined as $[\mathcal{A}(\mathbf{k})]_{mn} = \langle \Psi_m(\mathbf{k})|\partial_{\mathbf{k}}|\Psi_n(\mathbf{k})\rangle$. The orthonormality leads to $\mathcal{A}_{mn}(\mathbf{k}) = -\mathcal{A}_{nm}^*(\mathbf{k})$, i.e., $\mathcal{A}(\mathbf{k})$ is anti-Hermitian, which guarantees that $\mathcal{W}_{\mathcal{P}}$ is a unitary matrix. Under a unitary basis transformation $|\Psi_n(\mathbf{k})\rangle \to U(\mathbf{k})|\Psi_n(\mathbf{k})\rangle \equiv |\Psi'_n(\mathbf{k})\rangle$, the Berry connection and Wilson line transform as

$$\mathcal{A}(\mathbf{k}) \to \mathcal{A}'(\mathbf{k}) = U^{\dagger}(\mathbf{k})\mathcal{A}(\mathbf{k})U(\mathbf{k}) + U^{\dagger}(\mathbf{k})\partial_{\mathbf{k}}U(\mathbf{k}), \quad \mathcal{W}_{\mathcal{P}} \to \mathcal{W}'_{\mathcal{P}} = U^{\dagger}(\mathbf{k}_f)\mathcal{W}_{\mathcal{P}}U(\mathbf{k}_i). \tag{2}$$

We observe that $\mathcal{A}(\mathbf{k})$ and $W_{\mathcal{P}}$ resemble the gauge potential and Wilson line in $U(N)$ gauge theory respectively, and $U(\mathbf{k})$ resembles a gauge transformation. Note that the transformation $U(\mathbf{k})$ should preserve eigenenergy, so it is arbitrary only if the $N$ bands are fully degenerate along the path $\mathcal{P}$. Considering a closed loop $\mathcal{L}: \mathbf{k}_0 \leftarrow \mathbf{k}_0$, with $\mathbf{k}_0$ being a base point, we can construct the following gauge invariant Wilson loop observable

$$e^{i\gamma_{\mathcal{L}}} = \det(W_{\mathcal{L}}). \tag{3}$$

Note that gauge transformations (2) on $\mathcal{W}_{\mathcal{L}}$ cancel out for a closed loop $\mathcal{L}$ after taking a determinant. The phase factor $e^{i\gamma_{\mathcal{L}}}$ is also independent of the base point $\mathbf{k}_0$ of $\mathcal{L}$.

Instead of a single set of eigenvectors at each $\mathbf{k}$, a non-Hermitian system carries two sets of eigenvectors, the left and right eigenvectors, $\{|\Psi_{L,m}(\mathbf{k})\rangle\}$ and $\{|\Psi_{R,n}(\mathbf{k})\rangle\}$, respectively, which are not the same in general. We take the left and right eigenvectors to be biorthonormal, $\langle \Psi_{L,m}(\mathbf{k})|\Psi_{R,n}(\mathbf{k})\rangle = \langle \Psi_{R,n}(\mathbf{k})|\Psi_{L,m}(\mathbf{k})\rangle = \delta_{mn}$. We can define two types of Wilson lines:

$$\mathcal{W}_{\mathcal{P}}^{LR} = \overline{\exp}\left[-\int_{\mathbf{k}_i}^{\mathbf{k}_f} d\mathbf{k}\mathcal{A}^{LR}(\mathbf{k})\right], \quad \mathcal{W}_{\mathcal{P}}^{RL} = \overline{\exp}\left[-\int_{\mathbf{k}_i}^{\mathbf{k}_f} d\mathbf{k}\mathcal{A}^{RL}(\mathbf{k})\right], \tag{4}$$

where the Berry connection $\mathcal{A}^{\alpha\beta}(\mathbf{k})$ is defined as $\mathcal{A}_{mn}^{\alpha\beta}(\mathbf{k}) = \langle \Psi_{\alpha,m}(\mathbf{k})|\partial_{\mathbf{k}}|\Psi_{\beta,n}(\mathbf{k})\rangle$. Under a basis transformation $|\Psi_{L,n}(\mathbf{k})\rangle \to U(\mathbf{k})|\Psi_{L,n}(\mathbf{k})\rangle$ and $|\Psi_{R,n}(\mathbf{k})\rangle \to U(\mathbf{k})|\Psi_{R,n}(\mathbf{k})\rangle$, which preserves the biorthonomality, the transformations on $\mathcal{A}^{\alpha\beta}(\mathbf{k})$ and $\mathcal{W}_{\mathcal{P}}^{\alpha\beta}$ are the same as in (2). Unlike in the Hermitian case, $\mathcal{W}_{\mathcal{P}}$ is not unitary in general. The biorthonormality of the left and right eigenvectors gives $\mathcal{A}_{mn}^{RL} = -(\mathcal{A}_{nm}^{LR})^*$, with which one can show the following relations

$$\mathcal{W}_{\overline{\mathcal{P}}}^{\alpha\beta} = \left(\mathcal{W}_{\mathcal{P}}^{\beta\alpha}\right)^{\dagger}, \quad \mathcal{W}_{\mathcal{P}}^{\alpha\beta}\left(\mathcal{W}_{\mathcal{P}}^{\beta\alpha}\right)^{\dagger} = \mathcal{W}_{\mathcal{P}}^{\alpha\beta}\mathcal{W}_{\overline{\mathcal{P}}}^{\alpha\beta} = \mathbb{1}, \tag{5}$$

where $\alpha \neq \beta$, $\bar{\mathcal{P}}$ is the inverse path of $\mathcal{P}$, and $\mathbb{1}$ is the $N \times N$ identity matrix. If one considers a closed loop $\mathcal{L}$, then we have

$$\det\left(\mathcal{W}_{\mathcal{L}}^{LR}\right) = e^{a_{\mathcal{L}}+i\gamma_{\mathcal{L}}}, \quad \det\left(\mathcal{W}_{\bar{\mathcal{L}}}^{LR}\right) = e^{-a_{\mathcal{L}}-i\gamma_{\mathcal{L}}}, \quad \det\left(\mathcal{W}_{\mathcal{L}}^{RL}\right) = e^{-a_{\mathcal{L}}+i\gamma_{\mathcal{L}}}, \tag{6}$$

where both $a_L$ and $\gamma_{\mathcal{L}}$ are real numbers. We emphasize that while $\mathcal{W}_{\mathcal{L}}^{\alpha\beta}$ is not unitary, the gauge transformation $U(\mathbf{k})$ is still unitary. Shifting the phase $\gamma_{\mathcal{L}}$ by multiples of $2\pi$ corresponds to performing a large gauge transformation $U(\mathbf{k})$ on the gauge potential $\mathcal{A}(\mathbf{k})$. Here, "large" corresponds to a nontrivial winding number from the loop $\mathcal{L}$ to $U(N)$ matrices [more precisely, the diagonal $U(1)$].

The field strength is defined as $\mathcal{F}_{\mu\nu}^{\alpha\beta} = \partial_\mu \mathcal{A}_\nu^{\alpha\beta} - \partial_\nu \mathcal{A}_\mu^{\alpha\beta} + [\mathcal{A}_\mu^{\alpha\beta}, \mathcal{A}_\nu^{\alpha\beta}]$, where $\mu,\nu$ are components of momentum $\mathbf{k}$. The Berry curvature (or magnetic field) is defined as $\mathcal{B}^{\alpha\beta} = \frac{i}{2} \sum_{\nu\lambda} \epsilon^{\mu\nu\lambda} \mathcal{F}_{\mu\nu}^{\alpha\beta}$, where $\epsilon^{\mu\nu\lambda}$ is the fully anti-symmetric Levi-Civita tensor. Equivalently, $\mathcal{B}^{\alpha\beta} = i\nabla \times \mathcal{A}^{\alpha\beta} + i\mathcal{A}^{\alpha\beta} \times \mathcal{A}^{\alpha\beta}$ for 3D systems. We have the following relation

$$e^{a_{\mathcal{L}}+i\gamma_{\mathcal{L}}} = \det(\mathcal{W}_{\mathcal{L}}^{LR}) = \exp\left[i \int_{\Sigma} d\mathbf{S} \cdot \mathrm{tr}\left(\mathcal{B}^{LR}\right)\right], \tag{7}$$

where $\Sigma$ is a surface with the boundary $\partial\Sigma = \mathcal{L}$, and taking determinant makes the path order irrelevant. This relation says that $\det(\mathcal{W}_{\mathcal{L}}^{LR})$ measures the total magnetic flux through the surface $\Sigma$. For Eq. (7) to hold, it requires $\mathcal{A}^{LR}$ to be well defined everywhere on the surface $\Sigma$. Note that $\mathrm{tr}\left(\mathcal{B}^{LR}\right)$ is complex in general. For a boundaryless $\Sigma$, the Chern number can be defined as

$$C_\Sigma = \frac{1}{2\pi} \Re \int_{\Sigma} d\mathbf{S} \cdot \mathrm{tr}\left(\mathcal{B}^{LR}\right). \tag{8}$$

The Berry curvature $\mathcal{B}^{RL}$ will give rise to the same $C_\Sigma$. Like the Hermitian case, $C_\Sigma$ must be integer. It follows from Eq. (7), that $\gamma_{\mathcal{L}}$ is defined up to multiples of $2\pi$, and that for boundaryless surface $\gamma_{\partial\Sigma} = \gamma_{\emptyset} = 0$. Whether the imaginary part of the Berry curvature contains any topological information is unclear to us at this moment.

It is useful to have a discrete expression of $\mathcal{W}_{\mathcal{P}}^{\alpha\beta}$ for numerical purpose. Let us discretize the path $\mathcal{P}: \mathbf{k}_f \leftarrow \mathbf{k}_i$ into $M$ small segments, with the middle points being $\mathbf{k}_1, \mathbf{k}_2, \ldots, \mathbf{k}_{M-1}$ and $\Delta \mathbf{k}_s = \mathbf{k}_{s+1} - \mathbf{k}_s$ being infinitesimal. We take $\mathbf{k}_0 \equiv \mathbf{k}_i$ and $\mathbf{k}_M \equiv \mathbf{k}_f$. Then, the Wilson line can be computed as

$$\begin{aligned}\mathcal{W}_{\mathcal{P}}^{\alpha\beta} &= \lim_{M\to\infty} e^{-\Delta\mathbf{k}_{M-1}\,\mathcal{A}^{\alpha\beta}(\mathbf{k}_{M-1})} \ldots e^{-\Delta\mathbf{k}_1\,\mathcal{A}^{\alpha\beta}(\mathbf{k}_1)} e^{-\Delta\mathbf{k}_0\,\mathcal{A}^{\alpha\beta}(\mathbf{k}_0)} \\ &= \lim_{M\to\infty} \prod_{s=0}^{M-1} \left[\mathbb{1} - \Delta\mathbf{k}_s\,\mathcal{A}^{\alpha\beta}(\mathbf{k}_s)\right] \\ &\equiv \lim_{M\to\infty} \prod_{s=0}^{M-1} \mathcal{W}_{\mathbf{k}_{s+1}\leftarrow\mathbf{k}_s}^{\alpha\beta}\end{aligned} \tag{9}$$

where, in the second line, we have used the approximation $\exp\left[-\Delta\mathbf{k}\,\mathcal{A}^{\alpha\beta}(\mathbf{k})\right] = \mathbb{1} - \mathcal{A}^{\alpha\beta}(\mathbf{k})\Delta\mathbf{k} + \mathcal{O}(\Delta\mathbf{k}^2)$. For the infinitesimal segment $\mathbf{k}_{s+1} \leftarrow \mathbf{k}_s$, the Wilson line can be expressed as

$$\begin{aligned}\left[\mathcal{W}_{\mathbf{k}_{s+1}\leftarrow\mathbf{k}_s}^{\alpha\beta}\right]_{mn} &= \delta_{mn} - \Delta\mathbf{k}_s\,\mathcal{A}_{mn}^{\alpha\beta}(\mathbf{k}_s) \\ &= \delta_{mn} - \Delta\mathbf{k}_s\,\langle\Psi_{\alpha,m}(\mathbf{k}_{s+1})|\partial_\mathbf{k}|\Psi_{\beta,n}(\mathbf{k}_s)\rangle \\ &= \delta_{mn} - \langle\Psi_{\alpha,m}(\mathbf{k}_{s+1})|(|\Psi_{\beta,n}(\mathbf{k}_{s+1})\rangle - |\Psi_{\beta,n}(\mathbf{k}_s)\rangle) \\ &= \langle\Psi_{\alpha,m}(\mathbf{k}_{s+1})|\Psi_{\beta,n}(\mathbf{k}_s)\rangle.\end{aligned} \tag{10}$$

where the biorthonormal condition is used to achieve the last line. This is a numerically friendly expression.

## II. WILSON LOOP AND CHERN NUMBER WITH NON-HERMITIAN SYMMETRY

Here we discuss Wilson loops and Chern number in the presence of a unitary non-Hermitian spatial symmetry $\mathcal{G}$, which maps momentum $\mathbf{k}$ to $g\mathbf{k}$. Again, consider $N$ bands that are separated from other bands on the paths, loops or surfaces under consideration. Let $\{|\Psi_{L,n}(\mathbf{k})\rangle\}$ and $\{|\Psi_{R,n}(\mathbf{k})\rangle\}$ be the biorthonormal left and right eigenvectors of $\mathcal{H}(\mathbf{k})$. As discussed in the main text, $\mathcal{G}|\Psi_{R,n}(\mathbf{k})\rangle$ and $\mathcal{G}|\Psi_{L,n}(\mathbf{k})\rangle$ are the left and right eigenstates of $\mathcal{H}(g\mathbf{k})$ with

eigenenergy being $E_{\tilde{n}}(g\,\mathbf{k}) = E_n(\mathbf{k})^*$, where $\tilde{n}$ is the corresponding band index. We note that the set of degenerate bands indexed by $\tilde{n}$ is not necessarily the same as those indexed by $n$. We expand $\mathcal{G}|\Psi_{R,n}(\mathbf{k})\rangle$ and $\mathcal{G}|\Psi_{L,n}(\mathbf{k})\rangle$ in the energy eigenstates $\{|\Psi_{L,\tilde{n}}(g\,\mathbf{k})\rangle\}$ and $\{|\Psi_{R,\tilde{n}}(g\,\mathbf{k})\rangle\}$:

$$\begin{aligned}\mathcal{G}|\Psi_{R,n}(\mathbf{k})\rangle &= \sum_{\tilde{n}} |\Psi_{L,\tilde{n}}(g\mathbf{k})\rangle \mathcal{S}_{g,R}^{\tilde{n}n}(\mathbf{k}), \\ \mathcal{G}|\Psi_{L,n}(\mathbf{k})\rangle &= \sum_{\tilde{n}} |\Psi_{R,\tilde{n}}(g\mathbf{k})\rangle \mathcal{S}_{g,L}^{\tilde{n}n}(\mathbf{k}).\end{aligned} \quad (11)$$

where the sewing matrices are given by $\mathcal{S}_{g,\alpha}^{\tilde{n}n}(\mathbf{k}) = \langle \Psi_{\alpha,\tilde{n}}(g\mathbf{k})|\mathcal{G}|\Psi_{\alpha,n}(\mathbf{k})\rangle$, with $\alpha = L, R$. Due to the unitarity of $\mathcal{G}$, $\mathcal{G}\mathcal{G}^\dagger = \mathcal{G}^\dagger\mathcal{G} = 1$, the sewing matrix $\mathcal{S}_{g,\alpha}(\mathbf{k})$ satisfies the following relation

$$\mathcal{S}_{g,\alpha}(\mathbf{k})\mathcal{S}_{g,\bar{\alpha}}^\dagger(\mathbf{k}) = \mathcal{S}_{g,\alpha}^\dagger(\mathbf{k})\mathcal{S}_{g,\bar{\alpha}}(\mathbf{k}) = \mathbb{1}, \quad (12)$$

for any $\alpha$ and $\mathbf{k}$, where we have introduced the notation $\bar{L} = R$, $\bar{R} = L$, and $\mathbb{1}$ is the $N \times N$ identity matrix.

To investigate how the non-Hermitian symmetry relates different Wilson lines, we use the discrete expression Eq. (9). The Wilson line on the infinitesimal segment $\mathbf{k}_{s+1} \leftarrow \mathbf{k}_s$ satisfies the following relation

$$\begin{aligned}\left[\mathcal{W}_{\mathbf{k}_{s+1}\leftarrow\mathbf{k}_s}^{\alpha\bar{\alpha}}\right]_{mn} &= \langle\Psi_{\alpha,m}(\mathbf{k}_{s+1})|\Psi_{\bar{\alpha},n}(\mathbf{k}_s)\rangle \\ &= \langle\Psi_{\alpha,m}(\mathbf{k}_{s+1})|\mathcal{G}^\dagger\mathcal{G}|\Psi_{\bar{\alpha},n}(\mathbf{k}_s)\rangle \\ &= \sum_{\tilde{m}\tilde{n}} \mathcal{S}_{g,\alpha}^{\tilde{m}m}(\mathbf{k}_{s+1})^* \langle\Psi_{\bar{\alpha},\tilde{m}}(g\mathbf{k}_{s+1})|\Psi_{\alpha,\tilde{n}}(g\mathbf{k}_s)\rangle \mathcal{S}_{g,\bar{\alpha}}^{\tilde{n}n}(\mathbf{k}_s) \\ &= \left[\mathcal{S}_{g,\alpha}^\dagger(\mathbf{k}_{s+1})\tilde{\mathcal{W}}_{g\mathbf{k}_{s+1}\leftarrow g\mathbf{k}_s}^{\bar{\alpha}\alpha}\mathcal{S}_{g,\bar{\alpha}}(\mathbf{k}_s)\right]_{mn},\end{aligned} \quad (13)$$

where $\tilde{\mathcal{W}}^{\alpha\bar{\alpha}}$ denotes the Wilson line associated with the bands indexed by $\tilde{n}$. It can be more compactly written as $\mathcal{W}_{\mathbf{k}_{s+1}\leftarrow\mathbf{k}_s}^{\alpha\bar{\alpha}} = \mathcal{S}_{g,\alpha}^\dagger(\mathbf{k}_{s+1})\tilde{\mathcal{W}}_{g\mathbf{k}_{s+1}\leftarrow g\mathbf{k}_s}^{\bar{\alpha}\alpha}\mathcal{S}_{g,\bar{\alpha}}(\mathbf{k}_s)$. Inserting this relation into (9) for a general path $\mathcal{P}: \mathbf{k}_f \leftarrow \mathbf{k}_i$ and using (12), we immediately obtain

$$\mathcal{W}_\mathcal{P}^{\alpha\bar{\alpha}} = \mathcal{S}_{g,\alpha}^\dagger(\mathbf{k}_f)\tilde{\mathcal{W}}_{g\mathcal{P}}^{\bar{\alpha}\alpha}\mathcal{S}_{g,\bar{\alpha}}(\mathbf{k}_i). \quad (14)$$

where $g\mathcal{P}: g\,\mathbf{k}_f \leftarrow g\,\mathbf{k}_i$ is the image of the path $\mathcal{P}$ under the action of $\mathcal{G}$.

Considering Wilson observables on a closed loop $\mathcal{L}: \mathbf{k}_0 \leftarrow \mathbf{k}_0$, we have

$$e^{a_\mathcal{L}+i\gamma_\mathcal{L}} = \det\left(\mathcal{W}_\mathcal{L}^{LR}\right) = \det\left(\mathcal{S}_{g,L}^\dagger(\mathbf{k}_0)\tilde{\mathcal{W}}_{g\mathcal{L}}^{RL}\mathcal{S}_{g,R}(\mathbf{k}_0)\right) = \det\left(\tilde{\mathcal{W}}_{g\mathcal{L}}^{RL}\right) = e^{-\tilde{a}_{g\mathcal{L}}+i\tilde{\gamma}_{g\mathcal{L}}}. \quad (15)$$

Accordingly,

$$\tilde{a}_{g\mathcal{L}} = -a_\mathcal{L}, \quad \tilde{\gamma}_{g\mathcal{L}} = \gamma_\mathcal{L}. \quad (16)$$

Note that the bands indexed by $n$ and $\tilde{n}$ may be the same (which is the case in the exceptional unconventional Weyl semimetal example), so are the quantities in the above relations. Taking the loop $\mathcal{L}$ to be infinitesimally small, we immediately have that the Berry curvature satisfies

$$\mathrm{tr}\left(\mathcal{B}^{\alpha\bar{\alpha}}(\mathbf{k})\right) = \sigma(g)\mathrm{tr}\left(\tilde{\mathcal{B}}^{\bar{\alpha}\alpha}(g\,\mathbf{k})\right)^*. \quad (17)$$

where $\sigma(g) = 1$ or $-1$ denotes if $\mathcal{G}$ is orientation-preserving or orientation-reversing. This further leads to that the Chern numbers satisfy $\tilde{C}_{g\Sigma} = C_\Sigma$, for any boundaryless surface $\Sigma$. If $\Sigma$ and $g\Sigma$ are homotopy equivalent up to orientation (i.e., can be continuously deformed to each other without sweeping through degenerate regions), we must have $\tilde{C}_{g\Sigma} = \sigma(g)C_\Sigma$. Combining the two relations, we have $C_\Sigma = 0$ if $\sigma(g) = -1$.

### III. RELATION BETWEEN $W(\mathbf{k}_{\mathrm{EP}})$ AND $W(g\mathbf{k}_{\mathrm{EP}})$

In practice, the exceptional manifold can be found by studying the characteristic polynomial $\chi_\mathbf{k}(E) = \det[\mathcal{H}(\mathbf{k})-E]$. Without additional symmetry-protected degeneracy, $\chi_{\mathbf{k}_{\mathrm{EP}}}(E) \propto [E(\mathbf{k}_{\mathrm{EP}}) - E]^\mu$ near an EP at $\mathbf{k}_{\mathrm{EP}}$, where the order $\mu$ is equal to the size of Jordan block. Accordingly, at $E = E(\mathbf{k}_{\mathrm{EP}})$,

$$\chi_{\mathbf{k}_{\mathrm{EP}}}(E) = \frac{\partial \chi_{\mathbf{k}_{\mathrm{EP}}}(E)}{\partial E} = \cdots = \frac{\partial^{\mu-1}\chi_{\mathbf{k}_{\mathrm{EP}}}(E)}{\partial E^{\mu-1}} = 0. \quad (18)$$

Considering both $\chi$ and $E$ are complex, we have $2\mu$ equations and $D+2$ variables for $D$-dimensional systems. For order-2 EPs in 3D systems, the number of variables is one more than that of equations, indicating that EPs generally form exceptional lines(ELs).

Here we prove the relation Eq. (12) in the main text for topological invariants of EPs on ELs. First consider the case that $\mathcal{G}$ is a non-Hermitian symmetry. According to the definition (11), the winding numbers $W(\mathbf{k}_{\rm EP})$ and $W(g\mathbf{k}_{\rm EP})$ are given by

$$W(\mathbf{k}_{\rm EP}) = \oint_{S_a^1} \frac{d\mathbf{k}}{2\pi i} \cdot \nabla_{\mathbf{k}} \log\det\left[\mathcal{H}(\mathbf{k}) - E(\mathbf{k}_{\rm EP})\right], \tag{19}$$

$$W(g\mathbf{k}_{\rm EP}) = \oint_{S_b^1} \frac{d\mathbf{k}}{2\pi i} \cdot \nabla_{\mathbf{k}} \log\det\left[\mathcal{H}(\mathbf{k}) - E(g\mathbf{k}_{\rm EP})\right], \tag{20}$$

where $S_a^1$ is a loop encircling the exceptional line at $\mathbf{k}_{\rm EP}$ and $S_b^1$ is a loop encircling the exceptional line at $g\mathbf{k}_{\rm EP}$. Let us perform a change of variable $\mathbf{k} = g\mathbf{k}'$ for the integral (20). Then,

$$\begin{aligned}
W(g\mathbf{k}_{\rm EP}) &= \oint_{S_c^1} \frac{d(g\mathbf{k}')}{2\pi i} \cdot \nabla_{g\mathbf{k}'} \log\det\left[\mathcal{H}(g\mathbf{k}') - E(g\mathbf{k}_{\rm EP})\right] \\
&= \oint_{S_c^1} \frac{d\mathbf{k}'}{2\pi i} \cdot \nabla_{\mathbf{k}'} \log\det\left[\mathcal{G}\mathcal{H}^\dagger(\mathbf{k}')\mathcal{G}^{-1} - E^*(\mathbf{k}_{\rm EP})\right] \\
&= \oint_{S_c^1} \frac{d\mathbf{k}'}{2\pi i} \cdot \nabla_{\mathbf{k}'} \log\det\left[\mathcal{H}^\dagger(\mathbf{k}') - E^*(\mathbf{k}_{\rm EP})\right] \\
&= -\sigma(S_a^1, S_c^1) W(\mathbf{k}_{\rm EP}),
\end{aligned} \tag{21}$$

where $S_c^1$ is a loop encircling the exceptional line at $\mathbf{k}_{\rm EP}$ such that it is mapped to $S_b^1$ under the action of the spatial symmetry $\mathcal{G}$. In the second line, we have used Eq. (2) in the main text for non-Hermtian spatial symmetry and the relation $E(g\mathbf{k}_{\rm EP}) = E^*(\mathbf{k}_{\rm EP})$. The transformation matrix $g$ in $d(g\mathbf{k}')$ cancels that in $\nabla_{g\mathbf{k}'}$. In the last line, the minus sign follows from the Hermitian conjugation. The factor $\sigma(S_a^1, S_c^1) = 1$ if $S_a^1$ and $S_c^1$ have the same orientation, and $\sigma(S_a^1, S_c^1) = -1$ if they have opposite orientations. In our convention, the orientation of the integral loops is induced by the orientation of the exceptional lines through the right-hand rule. Then, it is not hard to show that

$$\sigma(S_a^1, S_c^1) = \sigma(g\mathbf{k}_{\rm EP}, \mathbf{k}_{\rm EP})\sigma(g). \tag{22}$$

The quantity $\sigma(g\mathbf{k}_{\rm EP}, \mathbf{k}_{\rm EP}) = 1$ or $-1$, corresponds to if the orientations of the two exceptional lines at $g\mathbf{k}_{\rm EP}$ and $\mathbf{k}_{\rm EP}$ agree or disagree under the symmetry action, respectively. The quantity $\sigma(g) = 1$ or $-1$, corresponds to if the spatial symmetry $\mathcal{G}$ preserves or reverses the global orientation of the momentum space (i.e., right-handed or left-handed).

The proof for $\mathcal{G}$ being a Hermitian symmetry is similar. The only difference is that there is no minus sign in the last line of (21).

## IV. SYMMETRIES IN EXCEPTIONAL UNCONVENTIONAL WEYL SEMIMETALS

In this section, we discuss the symmetries and symmetry-allowed non-Hermitian terms in the example of exceptional unconventional Weyl semimetals. Without the non-Hermitian term, the Hamiltonian reads

$$\mathcal{H}(\mathbf{k}) = 2A(\cos k_x - \cos k_y)\sigma_1 + 2A\sin k_x \sin k_y \sigma_2 + [M(k_x, k_y) - 2t_z \cos k_z]\sigma_3, \tag{23}$$

where $M(k_x, k_y) = M_0 - 2t_\parallel(\cos k_x + \cos k_y)$. The Hamiltonian respects the fourfold rotation symmetry about the $k_z$ axis,

$$\mathcal{C}_{4z}\mathcal{H}(\mathbf{k})\mathcal{C}_{4z}^{-1} = \mathcal{H}(R_{4z}\mathbf{k}): \quad \mathcal{C}_{4z} = \sigma_3, \text{ and } R_{4z}(k_x, k_y, k_z) = (k_y, -k_x, k_z). \tag{24}$$

We will analyze all symmetry-allowed constant perturbations to $\mathcal{H}(\mathbf{k})$. For two-band models, it suffices to employ Pauli matrices for discussion. Under the $\mathcal{C}_{4z}$ rotation operation, the Pauli matrices transform as

$$\mathcal{C}_{4z}\sigma_1\mathcal{C}_{4z}^{-1} = -\sigma_1, \quad \mathcal{C}_{4z}\sigma_2\mathcal{C}_{4z}^{-1} = -\sigma_2, \quad \mathcal{C}_{4z}\sigma_3\mathcal{C}_{4z}^{-1} = +\sigma_3. \tag{25}$$

It can be seen $\sigma_3$ is the only Hermitian symmetry-allowed term. A Hermitian $\lambda\sigma_3$ term can be absorbed by $M(k_x, k_y)$ in the model Hamiltonian of Eq. (23).

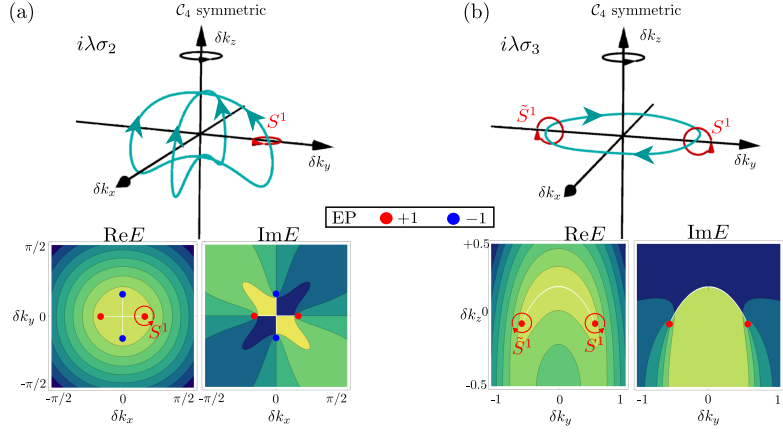

Fig. 1. (a) The exceptional structure from the non-Hermitian $i\lambda\sigma_2$, which respects the non-Hermitian $\mathcal{C}_{4z}$ symmetry. It is equivalent to $i\lambda\sigma_1$ discussed in the main text. The lower panel shows the contour plots of the real and imaginary eigenenergies at $\delta k_z = 0$. (b) The non-Hermitian term $i\lambda\sigma_3$ turns the unconventional Weyl points to exceptional rings. This term breaks the non-Hermitian $\mathcal{C}_{4z}$ symmetry, but it respects the Hermitian $\mathcal{C}_{4z}$ symmetry. The lower panel shows the contour plots at $\delta k_x = 0$. The parameters are the same as the main text.

Next, we focus on non-Hermitian symmetry-allowed constant terms denoted by $i\lambda\sigma$. Here $\lambda$ is real. The non-Hermitian $\mathcal{C}_{4z}$ symmetry allowed terms are

$$\mathcal{C}_{4z}(i\lambda\sigma)\mathcal{C}_{4z}^{-1} = (i\lambda\sigma)^\dagger: \qquad \sigma = \sigma_1,\ \sigma_2. \tag{26}$$

Thus, apart from $i\lambda\sigma_1$ discussed in the main text, $i\lambda\sigma_2$ is also a non-Hermitian $\mathcal{C}_{4z}$ symmetry allowed non-Hermitian term. As shown in Fig. 1(a), this term has the same effect as $i\lambda\sigma_1$ in the main text.

It should be noted that even though the non-Hermitian term $i\lambda\sigma_3$ does not respect non-Hermitian $\mathcal{C}_{4z}$ symmetry, it respects the Hermitian $\mathcal{C}_{4z}$ symmetry. The symmetry relation is given by

$$\mathcal{C}_{4z}(i\lambda\sigma_3)\mathcal{C}_{4z}^{-1} \neq (i\lambda\sigma_3)^\dagger, \quad \mathcal{C}_{4z}(i\lambda\sigma_3)\mathcal{C}_{4z}^{-1} = (i\lambda\sigma_3). \tag{27}$$

As shown in Fig. 1(b), this non-Hermitian term turns a Weyl point into an exceptional ring. It can also be explained by the effective theory around the original Weyl points. After the inclusion of $i\lambda\sigma_3$, the effective Hamiltonian becomes

$$\mathcal{H}_{\text{eff}}(\mathbf{K}_+ + \delta\mathbf{k}) \approx (\delta k_y^2 - \delta k_x^2)\sigma_1 + 2\delta k_x \delta k_y \sigma_2 + \delta k_z \sigma_3 + i\lambda\sigma_3. \tag{28}$$

Around the original Weyl point, i.e., $\delta k_z = 0$, the exceptional structure is determined by $(\delta k_y^2 - \delta k_x^2)^2 + (2\delta k_x \delta k_y)^2 - \lambda^2 = 0$. This yields $\delta k_x^2 + \delta k_y^2 = |\lambda|$, which is a ring. The exceptional structure is also symmetric with respect to $\mathcal{C}_{4z}$ symmetry, as shown in Fig. 1(b). This can be seen by the two exceptional points encircled by $\tilde{S}^1$ and $S^1$, their corresponding winding number satisfies the Hermitian symmetry relation.

We proceed to show in detail the transition from system with the non-Hermitian $\mathcal{C}_{4z}$ symmetry to that with Hermitian $\mathcal{C}_{4z}$, by introducing a $\delta H = i\lambda[(1-\eta)\sigma_1 + \eta\sigma_3]$ term,

$$\mathcal{H}^{w'}(\mathbf{k}) = \mathcal{H}_0^w(\mathbf{k}) + i\lambda[(1-\eta)\sigma_1 + \eta\sigma_3]. \tag{29}$$

At $\eta = 0$, the system respects the non-Hermitian $\mathcal{C}_{4z}$ symmetry, while at $\eta = 1$, the system respects the Hermitian $\mathcal{C}_4$ symmetry. In between, i.e., $0 < \eta < 1$, both Hermitian and non-Hermitian symmetries are broken. The evolution of exceptional manifolds against $\eta$ is shown in Fig. 2 upper panel. Clearly, for both Hermitian and non-Hermitian symmetries, the exceptional structure is $\mathcal{C}_{4z}$-symmetric. While in between, the exceptional structure is not $\mathcal{C}_{4z}$-symmetric.

The exceptional manifold at $\eta = 0$ is explicitly protected by the non-Hermitian $\mathcal{C}_{4z}$ symmetry. This can be seen by introducing a small $\eta$ to break the symmetry. To be concrete, we focus on how the exceptional intersection point $\mathbf{K}_j$, where two ELs join, as shown by red circle in Fig. 2 upper panel, is broken by the perturbation. The perturbed Hamiltonian around the intersection point $\mathbf{K}_j = (0, 0, \lambda)$ can be approximated by

$$\mathcal{H}_{\text{eff}}(\mathbf{K}_j + \delta\mathbf{k}) \approx (\delta k_y^2 - \delta k_x^2)\sigma_1 + 2\delta k_x \delta k_y \sigma_2 + (\lambda + \delta k_z)\sigma_3 + i\lambda\sigma_1 + i\varepsilon\sigma_3. \tag{30}$$

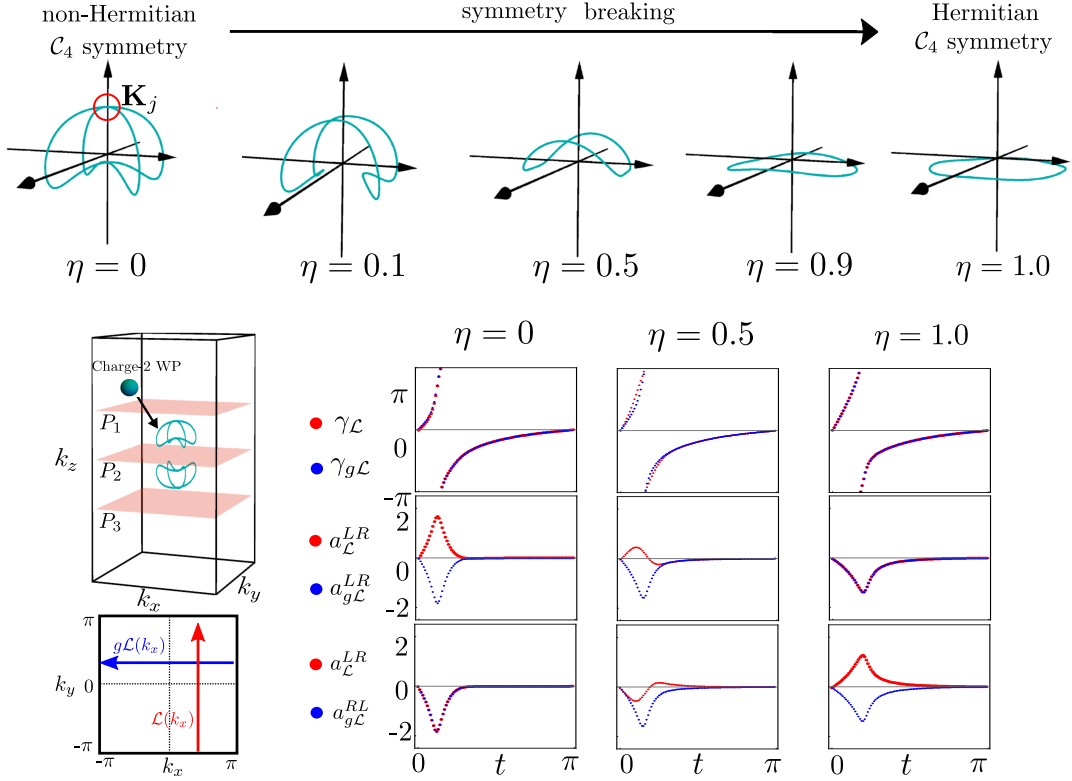

Fig. 2. Upper: The detail of evolution of exceptional manifold against $\eta$, in the transition between non-Hermitian $\mathcal{C}_{4z}$ symmetry and Hermitian $\mathcal{C}_{4z}$ symmetry. Lower: The loops $\mathcal{L}$ and $g\mathcal{L}$ refer to the red and blue arrows on plane $P_2$ in the first column. The right three columns show the transformation rules of Wilson loops on the paths $\mathcal{L}$ and $g\mathcal{L}$ under Hermitian and non-Hermitian spatial symmetries.

Here $\varepsilon = 0^+$. Using the characteristic polynomial, the exceptional manifold is determined by

$$\begin{aligned}(\delta k_x^2 + \delta k_y^2)^2 + (\delta k_z + \lambda)^2 - (\varepsilon^2 + \lambda^2) &= 0, \\ \lambda(\delta k_x^2 - \delta k_y^2) - \varepsilon(\delta k_z + \lambda) &= 0.\end{aligned} \quad (31)$$

We find that the symmetry breaking term $\varepsilon$, if non-vanishing, breaks the intersection point $\mathbf{K}_j$ joined by ELs. This can be seen by the second equation above. For $\varepsilon = 0$, when the non-Hermitian symmetry is respected, on the $k_z$ axis, $\delta k_x^2 - \delta k_y^2 = 0$ yields just one point, i.e., $\delta k_x = \delta k_y = 0$, which respects the symmetry. However, for non-zero $\varepsilon$, on the $k_z$ axis, $\delta k_x = \delta k_y = 0$ as required by the $\mathcal{C}_{4z}$ symmetry can no longer be achieved any more. Thus, the ELs can not join the intersection point on the $k_z$ axis.

Finally, we also visualize the difference in Wilson loops for non-Hermitian and Hermitian symmetries in Fig. 2 lower panel. We consider a family of non-contractible loops $\mathcal{L}: (t, -\pi, 0) \to (t, \pi, 0)$ (red arrow in the first column of lower panel) with $t \in [0, \pi]$. Under $\mathcal{C}_{4z}$, $\mathcal{L}$ is mapped to $g\mathcal{L}: (\pi, t, 0) \to (-\pi, t, 0)$ (blue arrow in the first column of lower panel). The quantities $a_{\mathcal{L}}^{\alpha\bar{\alpha}}$, $\gamma_{\mathcal{L}}^{\alpha\bar{\alpha}}$, $a_{g\mathcal{L}}^{\alpha\bar{\alpha}}$ and $\gamma_{g\mathcal{L}}^{\alpha\bar{\alpha}}$ are plotted in Fig. 2 (right three columns of lower panel) against the parameter $t$, which agrees with Table I in the main text ($\mathcal{C}_{4z}$ maps each band to itself in our model, so all quantities in Table I are associated with the same band).

## V. MULTIPLE SYMMETRIES AND HIGHER-ORDER EPS IN EXCEPTIONAL TRIPLE-POINT SEMIMETALS

Here we discuss a model that involves multiple non-Hermitian spatial symmetries and the appearance of higher-order EPs, i.e., multifold degeneracy with more than two eigenvectors coalescing to one. It is based on a Hermitian

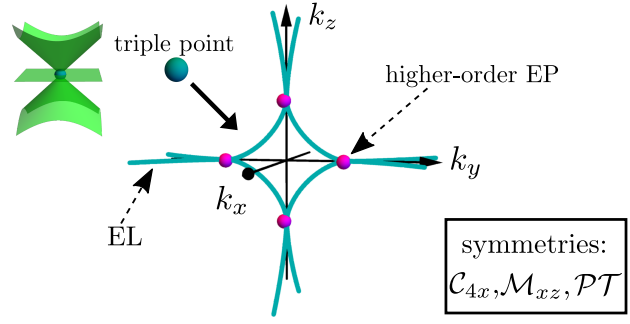

Fig. 3. Exceptional manifold of the exceptional triple-point semimetal. The parameters are $\alpha = 1$, $\beta = 1$, and $\lambda = 0.5$.

triple-point semimetal [1, 2], with the low-energy momentum-space Hamiltonian given by

$$\mathcal{H}^t(\mathbf{k}) = \mathcal{H}_0^t(\mathbf{k}) + i\lambda\Lambda$$
$$= \begin{pmatrix} \alpha k_x & \beta k_z & -\beta k_y \\ \beta k_z & -\alpha k_x & 0 \\ -\beta k_y & 0 & -\alpha k_x \end{pmatrix} + i\lambda \begin{pmatrix} 0 & 0 & 0 \\ 0 & 0 & 1 \\ 0 & 1 & 0 \end{pmatrix}. \tag{32}$$

Without the non-Hermitian term $i\lambda\Lambda$, the triple point is stabilized by three symmetries: a fourfold rotation symmetry about the $x$ axis ($\mathcal{C}_{4x}$), a mirror symmetry about $xz$-plane ($\mathcal{M}_{xz}$), and a combined symmetry of space-inversion and time-reversal symmetry ($\mathcal{PT}$), which read

$$\mathcal{C}_{4x}\mathcal{H}_0^t(\mathbf{k})\mathcal{C}_{4x}^{-1} = \mathcal{H}_0^t(R_{4x}\mathbf{k}), \tag{33}$$
$$\mathcal{M}_{xz}\mathcal{H}_0^t(\mathbf{k})\mathcal{M}_{xz}^{-1} = \mathcal{H}_0^t(M_{xz}\mathbf{k}), \tag{34}$$
$$\mathcal{PT}\mathcal{H}_0^t(\mathbf{k})(\mathcal{PT})^{-1} = \mathcal{H}_0^t(\mathbf{k}). \tag{35}$$

The symmetry operators are

$$\mathcal{C}_{4x} = \begin{pmatrix} -1 & 0 & 0 \\ 0 & 0 & 1 \\ 0 & -1 & 0 \end{pmatrix}, \qquad R_{4x}: (k_x, k_y, k_z) \to (k_x, -k_z, k_y), \tag{36}$$

$$\mathcal{M}_{xz} = \begin{pmatrix} 1 & 0 & 0 \\ 0 & 1 & 0 \\ 0 & 0 & -1 \end{pmatrix}, \qquad M_{xz}: (k_x, k_y, k_z) \to (k_x, -k_y, k_z), \tag{37}$$

$$\mathcal{PT} = \begin{pmatrix} 1 & 0 & 0 \\ 0 & 1 & 0 \\ 0 & 0 & 1 \end{pmatrix}\mathcal{K}, \qquad \mathcal{PT}: (k_x, k_y, k_z) \to (k_x, k_y, k_z). \tag{38}$$

Here $\mathcal{K}$ is the complex conjugation operation. Such a triply degenerate point can be found in materials of space group No. 225 [3].

The full non-Hermitian Hamiltonian $\mathcal{H}^t(\mathbf{k})$ respects $\mathcal{C}_{4x}$, $\mathcal{M}_{xz}$, and $\mathcal{PT}$ all as non-Hermitian symmetries,

$$\mathcal{C}_{4x}\mathcal{H}^t(\mathbf{k})\mathcal{C}_{4x}^{-1} = \mathcal{H}^t(R_{4x}\mathbf{k})^\dagger, \tag{39}$$
$$\mathcal{M}_{xz}\mathcal{H}^t(\mathbf{k})\mathcal{M}_{xz}^{-1} = \mathcal{H}^t(M_{xz}\mathbf{k})^\dagger, \tag{40}$$
$$\mathcal{PT}\mathcal{H}^t(\mathbf{k})(\mathcal{PT})^{-1} = \mathcal{H}^t(\mathbf{k})^\dagger. \tag{41}$$

As shown in Fig. 3, the non-Hermitian term $i\lambda\Lambda$ turns the triple point into an exceptional manifold, computed by the characteristic polynomial according to Eq. (18). We find that apart from order-2 ELs (green lines), there are four order-3 EPs (magenta points). Different from the exceptional Weyl system, the triple point does not have a well-defined topological invariant like Chern number, because a fully gapped sphere surrounding it cannot be found. The relation (12) in the main text is confirmed for both the orientation-preserving $\mathcal{C}_{4x}$ and orientation-reversing $\mathcal{M}_{xz}$ non-Hermitian spatial symmetries.

Next we show that the stability of the exceptional manifold is perturbatively protected by the non-Hermitian symmetries because $i\lambda\Lambda$ is the only symmetry-allowed term. In order to discuss symmetry-allowed terms, we employ the $3 \times 3$ Gell-Mann matrices in the following form,

$$\Lambda_1 = \begin{pmatrix} 0 & 1 & 0 \\ 1 & 0 & 0 \\ 0 & 0 & 0 \end{pmatrix}, \quad \Lambda_2 = \begin{pmatrix} 0 & -i & 0 \\ i & 0 & 0 \\ 0 & 0 & 0 \end{pmatrix}, \quad \Lambda_3 = \begin{pmatrix} 1 & 0 & 0 \\ 0 & -1 & 0 \\ 0 & 0 & 0 \end{pmatrix}, \quad \Lambda_4 = \begin{pmatrix} 0 & 0 & 1 \\ 0 & 0 & 0 \\ 1 & 0 & 0 \end{pmatrix}, \tag{42}$$

$$\Lambda_5 = \begin{pmatrix} 0 & 0 & -i \\ 0 & 0 & 0 \\ i & 0 & 0 \end{pmatrix}, \quad \Lambda_6 = \begin{pmatrix} 0 & 0 & 0 \\ 0 & 0 & 1 \\ 0 & 1 & 0 \end{pmatrix}, \quad \Lambda_7 = \begin{pmatrix} 0 & 0 & 0 \\ 0 & 0 & -i \\ 0 & i & 0 \end{pmatrix}, \quad \Lambda_8 = \begin{pmatrix} \frac{1}{\sqrt{3}} & 0 & 0 \\ 0 & \frac{1}{\sqrt{3}} & 0 \\ 0 & 0 & -\frac{2}{\sqrt{3}} \end{pmatrix}. \tag{43}$$

We investigate how the Gell-Mann matrices transform under each symmetry operation as follows.

**(1)** Under $\mathcal{C}_{4x}$ operation, the transformation of Gell-Mann matrices can be separated into 5 groups, which read

$$\begin{aligned} \{\mathcal{C}_{4x}\Lambda_1\mathcal{C}_{4x}^{-1} &= \Lambda_4, \quad \mathcal{C}_{4x}\Lambda_4\mathcal{C}_{4x}^{-1} = -\Lambda_1\}, \qquad \{\mathcal{C}_{4x}\Lambda_2\mathcal{C}_{4x}^{-1} = \Lambda_5, \quad \mathcal{C}_{4x}\Lambda_5\mathcal{C}_{4x}^{-1} = -\Lambda_2\}, \\ \{\mathcal{C}_{4x}\Lambda_3\mathcal{C}_{4x}^{-1} &= \frac{1}{2}(\Lambda_3 + \sqrt{3}\Lambda_8), \quad \mathcal{C}_{4x}\Lambda_8\mathcal{C}_{4x}^{-1} = \frac{1}{2}(\sqrt{3}\Lambda_3 - \Lambda_8)\}, \\ \{\mathcal{C}_{4x}\Lambda_6\mathcal{C}_{4x}^{-1} &= -\Lambda_6\}, \qquad \{\mathcal{C}_{4x}\Lambda_7\mathcal{C}_{4x}^{-1} = \Lambda_7\}. \end{aligned} \tag{44}$$

**(2)** Under $\mathcal{M}_{xz}$ operation, the Gell-Mann matrices transforms as

$$\begin{aligned} \{\mathcal{M}_{xz}\Lambda_1\mathcal{M}_{xz}^{-1} &= \Lambda_1, \quad \mathcal{M}_{xz}\Lambda_2\mathcal{M}_{xz}^{-1} = \Lambda_2, \quad \mathcal{M}_{xz}\Lambda_3\mathcal{M}_{xz}^{-1} = \Lambda_3, \quad \mathcal{M}_{xz}\Lambda_8\mathcal{M}_{xz}^{-1} = \Lambda_8\}, \\ \{\mathcal{M}_{xz}\Lambda_4\mathcal{M}_{xz}^{-1} &= -\Lambda_4, \quad \mathcal{M}_{xz}\Lambda_5\mathcal{M}_{xz}^{-1} = -\Lambda_5, \quad \mathcal{M}_{xz}\Lambda_6\mathcal{M}_{xz}^{-1} = -\Lambda_6, \quad \mathcal{M}_{xz}\Lambda_7\mathcal{M}_{xz}^{-1} = -\Lambda_7\}. \end{aligned} \tag{45}$$

**(3)** Under $\mathcal{PT}$ operation, the Gell-Mann matrices transforms as

$$\{\mathcal{PT}\Lambda_1(\mathcal{PT})^{-1} = \Lambda_1, \quad \mathcal{PT}\Lambda_3(\mathcal{PT})^{-1} = \Lambda_3, \quad \mathcal{PT}\Lambda_4(\mathcal{PT})^{-1} = \Lambda_4, \quad \mathcal{PT}\Lambda_6(\mathcal{PT})^{-1} = \Lambda_6, \quad \mathcal{PT}\Lambda_8(\mathcal{PT})^{-1} = \Lambda_8\},$$
$$\{\mathcal{PT}\Lambda_2(\mathcal{PT})^{-1} = -\Lambda_2, \quad \mathcal{PT}\Lambda_5(\mathcal{PT})^{-1} = -\Lambda_5, \quad \mathcal{PT}\Lambda_7(\mathcal{PT})^{-1} = -\Lambda_7\}. \tag{46}$$

**Hermitian symmetry-allowed terms**: Before discussing the non-Hermitian symmetries, we first investigate conventional Hermitian symmetries. We focus on the Hermitian constant terms denoted as $\lambda\Lambda$ that are invariant under the symmetry operation.

The $\mathcal{C}_{4x}$ symmetry-allowed terms are

$$\mathcal{C}_{4x}(\lambda\Lambda)(\mathbf{k})\mathcal{C}_{4x}^{-1} = (\lambda\Lambda): \qquad \Lambda = \left(\sqrt{3}\Lambda_3 + \Lambda_8\right), \Lambda_7. \tag{47}$$

The $\mathcal{M}_{xz}$ symmetry-allowed terms are

$$\mathcal{M}_{xz}(\lambda\Lambda)\mathcal{M}_{xz}^{-1} = (\lambda\Lambda): \qquad \Lambda = \Lambda_1, \Lambda_2, \Lambda_3, \Lambda_8. \tag{48}$$

The $\mathcal{PT}$ symmetry-allowed terms are

$$\mathcal{PT}(\lambda\Lambda)(\mathcal{PT})^{-1} = (\lambda\Lambda): \qquad \Lambda = \Lambda_1, \Lambda_3, \Lambda_4, \Lambda_6, \Lambda_8. \tag{49}$$

By comparing Eqs. (47), (48), and (49), we can find just one term that respects all the symmetries.

$$\Lambda = \left(\sqrt{3}\Lambda_3 + \Lambda_8\right). \tag{50}$$

Such a term, however, just shifts the location of the triple point on $k_x$ axis. This can be seen by expressing the Hamiltonian in terms of Gell-Mann Matrices,

$$\mathcal{H}_0(\mathbf{k}) = \begin{pmatrix} \alpha k_x & \beta k_z & -\beta k_y \\ \beta k_z & -\alpha k_x & 0 \\ -\beta k_y & 0 & -\alpha k_x \end{pmatrix} = -\frac{1}{3}\alpha k_x \mathcal{I} + \frac{\alpha}{\sqrt{3}} k_x \left(\sqrt{3}\Lambda_3 + \Lambda_8\right) - \beta k_y \Lambda_4 + \beta k_z \Lambda_1. \tag{51}$$

After the inclusion of symmetry-allowed term, the Hamiltonian becomes

$$\mathcal{H}_0(\mathbf{k}) + \lambda\Lambda = -\frac{1}{3}\alpha k_x \mathcal{I} + \frac{\alpha}{\sqrt{3}}(k_x + \lambda)\left(\sqrt{3}\Lambda_3 + \Lambda_8\right) - \beta k_y \Lambda_4 + \beta k_z \Lambda_1. \tag{52}$$

Thus, the Hermitian symmetry-allowed term just shifts the location of the triple point in momentum space.

**Non-Hermitian symmetry-allowed terms**: We now turn to the non-Hermitian spatial symmetries. Consider the symmetry-allowed non-Hermitian terms in the form of $(i\lambda\Lambda)$, with $\lambda$ real and $\Lambda$ Hermitian. Note that the following relation

$$(i\lambda\Lambda)^\dagger = -i\lambda\Lambda, \tag{53}$$

will be used together with symmetry operations in Eqs. (44), (45), and (46).

The non-Hermitian $\mathcal{C}_{4x}$ symmetry-allowed terms are

$$\mathcal{C}_{4x}(i\lambda\Lambda)\mathcal{C}_{4x}^{-1} = (i\lambda\Lambda)^\dagger : \qquad \Lambda = \left(\Lambda_3 - \sqrt{3}\Lambda_8\right), \Lambda_6. \tag{54}$$

The non-Hermitian $\mathcal{M}_{xz}$ symmetry-allowed terms are

$$\mathcal{M}_{xz}(i\lambda\Lambda)\mathcal{M}_{xz} = (i\lambda\Lambda)^\dagger : \qquad \Lambda = \Lambda_4, \Lambda_5, \Lambda_6, \Lambda_7. \tag{55}$$

The non-Hermitian $\mathcal{PT}$ symmetry-allowed terms are

$$(\mathcal{PT})(i\lambda\Lambda)(\mathcal{PT})^{-1} = (i\lambda\Lambda)^\dagger : \qquad \Lambda = \Lambda_1, \Lambda_3, \Lambda_4, \Lambda_6, \Lambda_8. \tag{56}$$

By comparing Eqs. (54), (55), and (56), there is only one non-Hermitian symmetry allowed term, namely,

$$\Lambda = \Lambda_6. \tag{57}$$

## VI. EXPERIMENTAL REALIZATION OF WEYL SEMIMETAL PROTECTED BY NON-HERMITIAN INVERSION SYMMETRY IN COLD ATOMS

Here we present the scheme details for realizing the model Hamiltonian (13) of Weyl semimetal protected by non-Hermitian inversion symmetry in the main text. We choose two atomic internal states as pseudo-spins, and load the atoms trapped in a two-dimensional optical lattice, as shown in Fig. 4. We assume the lattice trap potential is $V_L(\mathbf{r}) = \sum_{\eta=x,y,z} V_L \sin^2(k_L\eta)$ ($k_L = \pi/d$ and $d$ denotes the optical lattice constant). Since the model Hamiltonian contains four degrees of freedom, we can load the atoms into a bilayer optical lattice and denote the atomic operator as $\psi_{\alpha,\beta}$, where $\alpha = 1, 2$ stands for the layer indices along the $z$ direction and $\beta = \uparrow\downarrow$ for the pseudo-spins. In this way, the layer index can act as the extra two degrees of freedom beside the two pseudo-spins. By preparing a deep trap potential, the natural nearest-neighbor hopping can be neglected. We consider the Hamiltonian in the following three parts,

$$H = H_1 + H_2 + H_3. \tag{58}$$

The first part describes the coupling between the two pseudo-spins,

$$H_1 = \int d\mathbf{r} \sum_{\alpha=1,2} M_1(\mathbf{r})\psi_{\alpha,\uparrow}^\dagger(\mathbf{r})\psi_{\alpha,\downarrow}(\mathbf{r}) + H.c., \tag{59}$$

where the optical field mode is

$$M_1(\mathbf{r}) = M_1 \sin(k_L x)\cos(k_L y)\cos(k_L z) + iM_1 \cos(k_L x)\sin(k_L y)\cos(k_L z). \tag{60}$$

The second part describes the coupling between the two layer indices,

$$H_2 = \int d\mathbf{r} \sum_{\beta=\uparrow\downarrow} M_2(\mathbf{r})\psi_{1,\beta}^\dagger(\mathbf{r})\psi_{2,\beta}(\mathbf{r}) + H.c., \tag{61}$$

where the optical field mode is

$$M_2(\mathbf{r}) = M_2 \cos(k_L x)\cos(k_L y)\sin(k_L z) - M_2' \cos(2k_L x)\cos(2k_L y)\sin(2k_L z). \tag{62}$$

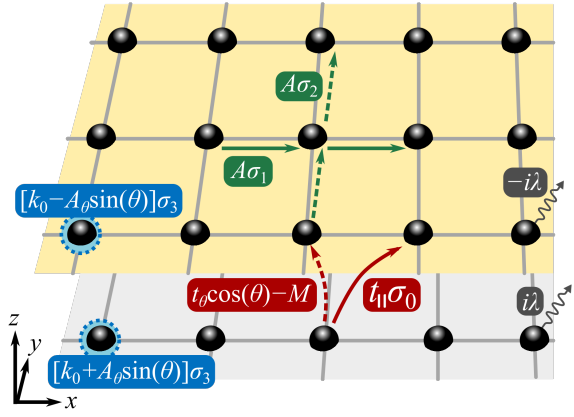

Fig. 4. Setups for Sec. VI. The $A$ (green lines) and $t_\theta$ (red lines) terms are generated via optical fields (60) and (62), respectively. The on-site energy $k_0 \tau_0 \sigma_3 + A_\theta \sin(\theta)\tau_3\sigma_3$ depends on the layer indices as well as pseudo-spins. $\pm i\lambda$ colored in gray represents the atomic decay, responsible for the non-Hermiticity. Notice that since the figure shows the two lattice layers aligned in the $z$ direction, the illustration of the third artificial dimension $\theta$ is therefore hidden.

The last part describes the on-site energy term as well as the atomic decay,

$$H_3 = \int d\mathbf{r} \sum_{\alpha,\beta} [\Gamma_\beta \cos(k_L z) + i\lambda \tau_3^{\alpha\alpha}] \psi_{\alpha,\beta}^\dagger(\mathbf{r}) \psi_{\alpha,\beta}(\mathbf{r}), \tag{63}$$

where $\Gamma_\beta$ stands for the energy of the pseudo-spin $\beta$. The decay depends on the layer indices, which can be introduced via preparing the number loss at different rates. By performing the same approach similar to that in Sec. VII, Eqs. (59), (61), and (63) give the tight-binding Hamiltonians as follows,

$$H_1 = \sum_{j,\alpha} \mathcal{M}_1 \tau_3^{\alpha\alpha} (c_{j+\hat{e}_x,\alpha,\uparrow}^\dagger c_{j,\alpha,\downarrow} - c_{j-\hat{e}_x,\alpha,\uparrow}^\dagger c_{j,\alpha,\downarrow}) + i\mathcal{M}_1 \tau_3^{\alpha\alpha} (c_{j+\hat{e}_y,\alpha,\uparrow}^\dagger c_{j,\alpha,\downarrow} - c_{j-\hat{e}_y,\alpha,\uparrow}^\dagger c_{j,\alpha,\downarrow}) + H.c. \tag{64}$$

$$H_2 = \sum_{j,\beta} \mathcal{M}_2 (c_{j+\hat{e}_x,1,\beta}^\dagger c_{j,2,\beta} + c_{j-\hat{e}_x,1,\beta}^\dagger c_{j,2,\beta} + c_{j+\hat{e}_y,1,\beta}^\dagger c_{j,2,\beta} + c_{j-\hat{e}_y,1,\beta}^\dagger c_{j,2,\beta}) + \mathcal{M}_2' c_{j,1,\beta}^\dagger c_{j,2,\beta} + H.c. \tag{65}$$

$$H_3 = \sum_j \sum_{\alpha,\alpha'} \sum_{\beta,\beta'} [k_0 \tau_0^{\beta\beta'} \sigma_3^{\alpha\alpha'} + A_\theta \sin(k_\theta) \tau_3^{\beta\beta'} \sigma_3^{\alpha\alpha'} + i\lambda \tau_3^{\beta\beta'} \sigma_0^{\alpha\alpha'}] c_{j,\beta,\alpha}^\dagger c_{j,\beta',\alpha'}. \tag{66}$$

Here

$$\mathcal{M}_1 = \int M_1 \sin(k_L x) \cos(k_L y) \cos(k_L z) W^*(\mathbf{r}+d\hat{e}_x) W(\mathbf{r}) d\mathbf{r}, \tag{67}$$

$$\mathcal{M}_2 = \int M_2 \cos(k_L x) \cos(k_L y) \sin(k_L z) W^*(\mathbf{r}+d\hat{e}_x + d\hat{e}_z) W(\mathbf{r}) d\mathbf{r}, \tag{68}$$

$$\mathcal{M}_2' = \int M_2(\mathbf{r}) W^*(\mathbf{r}+d\hat{e}_z) W(\mathbf{r}) d\mathbf{r}, \tag{69}$$

in which we have accounted for the homogeneity between the $x$ and $y$ directions, and the spatial symmetry of $W(\mathbf{r})$ with respect to the site center. We have also tuned the on-site energy with a tunable cyclical parameter $\theta$ [4],

$$(\Gamma_\uparrow + \Gamma_\downarrow)/2 = k_0, \qquad (\Gamma_\uparrow - \Gamma_\downarrow)/2 = A_\theta \sin(\theta). \tag{70}$$

By making the following denotations,

$$2\mathcal{M}_1 \equiv A, \qquad \mathcal{M}_2 \equiv t_\parallel, \qquad \mathcal{M}_2' \equiv t_\theta \cos(\theta) - M, \tag{71}$$

and the transformation into the momentum space, we obtain the Hamiltonian as

$$H(\mathbf{k}) = A \sin(k_x d)\tau_3 \sigma_1 + A \sin(k_y d)\tau_3 \sigma_2 + A_\theta \sin(k_\theta)\tau_3 \sigma_3 \\ + [t_\parallel \cos(k_x d) + t_\parallel \cos(k_y d) + t_\theta \cos(\theta) - M]\tau_1 \sigma_0 + k_0 \tau_0 \sigma_3 + i\lambda \tau_3 \sigma_0, \tag{72}$$

which is illustrated in Fig. 4. One can find that Eq. (72) is identical to the model Hamiltonian (13) in the main text, by recognizing $\theta$ as the artificial dimension $k_z$ and making $\tau_3 \leftrightarrow \tau_1$.

## VII. EXPERIMENTAL REALIZATION OF EXCEPTIONAL UNCONVENTIONAL WEYL SEMIMETAL IN COLD ATOMS

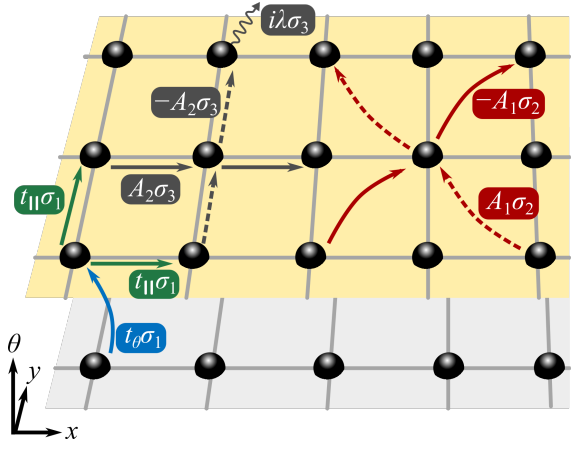

Fig. 5. Illustration of the setups for Sec. VII. We show the $x$-$y$ plane accompanied by the third artificial dimension $\theta$. The $A_1$ (red lines), $t_\parallel$ (green lines), and $t_\theta$ (blue lines) are generated via optical field modes $M_1(\mathbf{r})$, $M_2(\mathbf{r})$, and $M_3(\mathbf{r})$ of Eq. (75), respectively. The $A_2$ terms originate from the natural nearest-neighbor hopping by making the operator transformation (88). The $\lambda$ term is introduced by the on-site atomic decay.

Next, we present the scheme details for realizing the model Hamiltonian (16) of exceptional unconventional Weyl semimetals in the main text using cold atoms. As shown in Fig. 5, we load the atoms into a single-layer lattice, where the pseudo-spins stand for the two degrees of freedom in the model Hamiltonian. We consider the following Hamiltonian composed of three parts,

$$H = H_1 + H_2 + H_3 \,. \tag{73}$$

The first part of Hamiltonian (73) describes the coupling between pseudo-spins [5],

$$H_1 = \int d\mathbf{r} M(\mathbf{r}) \psi_\uparrow^\dagger(\mathbf{r}) \psi_\downarrow(\mathbf{r}) + H.c. \tag{74}$$

It is generated by three optical fields,

$$M(\mathbf{r}) = M_1(\mathbf{r}) + M_2(\mathbf{r}) + M_3(\mathbf{r}) \,, \tag{75}$$

whose spatial modulations are prepared as

$$M_1(\mathbf{r}) = iM_1 \sin(k_L x) \sin(k_L y) \,, \tag{76}$$
$$M_2(\mathbf{r}) = M_2[\cos(k_L x)\cos(3k_L y) - \cos(3k_L x)\cos(k_L y)] \,, \tag{77}$$
$$M_3(\mathbf{r}) = M_3 \cos(2k_L x)\cos(2k_L y) \,. \tag{78}$$

$M_{1,2,3}$ denotes the corresponding field strength.

We expand the operator $\psi_\nu(\mathbf{r})$ in terms of Wannier wave function $W(\mathbf{r})$:

$$\psi_\nu(\mathbf{r}) = \sum_j W(\mathbf{r} - \mathbf{r}_j) c_{j,\nu} \,. \tag{79}$$

Here $j$ stands for the site index, $W(\mathbf{r})$ is assumed to be localized and centered on $\mathbf{r} = 0$, and we denote the coordinate $\mathbf{r}_j = (j_x, j_y, j_z) \times d$. Under the transformation of Eq. (79), $H_1$ is rewritten as

$$H_1 = \sum_{\eta=1,2,3} \mathcal{M}_\eta^{(i,j)} c_{i,\uparrow}^\dagger c_{j,\downarrow} \,, \tag{80}$$

where $\mathcal{M}_\eta^{(i-j)}$ denotes the spin coupling between the $i$-th and $j$-th sites,

$$\mathcal{M}_\eta^{(i,j)} = \int M_\eta(\mathbf{r})W^*(\mathbf{r}-\mathbf{r}_i)W(\mathbf{r}-\mathbf{r}_j)d\mathbf{r}. \tag{81}$$

The field modes $M_\eta(\mathbf{r})$ give rise to the following results: (i) Due to the odd parity of Eq. (76) in the $x$-$y$ plane, the on-site $\mathcal{M}_1^{(j,j)}$ and nearest-neighbor coupling terms $\mathcal{M}_1^{(j\pm1,j)}$ vanish, leaving the next-nearest-neighbor term $\mathcal{M}_1^{(j\pm2,j)}$ dominant. (ii) As the system is homogeneous in the $x$-$y$ plane, the on-site coupling term $\mathcal{M}_2^{(j,j)}$ introduced by Eq. (77) vanishes [6], leaving the nearest-neighbor term $\mathcal{M}_2^{(j\pm1,j)}$ dominant. (iii) Eq. (78) does not introduce nonlocal coupling because of the spatial symmetry of $W(\mathbf{r})$ with respect to the site center.

Based on the aforementioned results, $H_1$ in the tight-binding approximation is expressed as

$$H_1 = \sum_j (-1)^{j_x+j_y} \Big[ i\mathcal{M}_1(-c^\dagger_{j+\hat{e}_x-\hat{e}_y\uparrow}c_{j,\downarrow} - c^\dagger_{j-\hat{e}_x+\hat{e}_y\uparrow}c_{j,\downarrow} + c^\dagger_{j+\hat{e}_x+\hat{e}_y\uparrow}c_{j,\downarrow} + c^\dagger_{j-\hat{e}_x-\hat{e}_y\uparrow}c_{j,\downarrow})$$
$$+ \mathcal{M}_2(c^\dagger_{j+\hat{e}_x\uparrow}c_{j,\downarrow} + c^\dagger_{j-\hat{e}_x\uparrow}c_{j,\downarrow} - c^\dagger_{j+\hat{e}_y\uparrow}c_{j,\downarrow} - c^\dagger_{j-\hat{e}_y\uparrow}c_{j,\downarrow}) + \mathcal{M}_3 c^\dagger_{j\uparrow}c_{j,\downarrow} \Big] + H.c. \tag{82}$$

where $\hat{e}_{x,y}$ denotes the unit vector and $j_x$ and $j_y$ respectively denote the $x$- and $y$-directional index of the $j$-th site, and

$$\mathcal{M}_1 = \int M_1 \sin(k_L x)\sin(k_L y) W^*(\mathbf{r}+d\hat{e}_x+d\hat{e}_y)W(\mathbf{r})d\mathbf{r}, \tag{83}$$

$$\mathcal{M}_2 = \int M_2(\mathbf{r}) W^*(\mathbf{r}+d\hat{e}_x)W(\mathbf{r})d\mathbf{r}, \tag{84}$$

$$\mathcal{M}_3 = \int M_3(\mathbf{r})|W(\mathbf{r})|^2 d\mathbf{r}. \tag{85}$$

The second part of Hamiltonian (73) describes the natural nearest-neighbor hopping of the tight-binding model,

$$H_2 = -\sum_{j,\nu} J(c^\dagger_{j+\hat{e}_x\nu}c_{j,\nu} + c^\dagger_{j+\hat{e}_y\nu}c_{j,\nu} + H.c.), \tag{86}$$

where $J$ is the hopping magnitude.

The last part of Hamiltonian (73) describes the on-site atomic decay,

$$H_3 = i\lambda \int [\psi^\dagger_\uparrow(\mathbf{r})\psi_\uparrow(\mathbf{r}) - \psi^\dagger_\downarrow(\mathbf{r})\psi_\downarrow(\mathbf{r})]d\mathbf{r}. \tag{87}$$

A recent advance in cold atoms [7] shows the experimental feasibility for engineering spin-dependent decays using the auxiliary level transition.

We make the following transformation,

$$c_{j\uparrow} \to (-1)^{j_x} c_{j\uparrow}, \qquad c_{j\downarrow} \to (-1)^{j_y} c_{j\downarrow}. \tag{88}$$

Under Eq. (88), the factor $(-1)^{j_x+j_y}$ in $H_1$ is eliminated and $H_1$ is rewritten as

$$H_1 = \sum_j \Big[ i\mathcal{M}_1(c^\dagger_{j+\hat{e}_x-\hat{e}_y\uparrow}c_{j,\downarrow} + c^\dagger_{j-\hat{e}_x+\hat{e}_y\uparrow}c_{j,\downarrow} - c^\dagger_{j+\hat{e}_x+\hat{e}_y\uparrow}c_{j,\downarrow} - c^\dagger_{j-\hat{e}_x-\hat{e}_y\uparrow}c_{j,\downarrow})$$
$$- \mathcal{M}_2(c^\dagger_{j+\hat{e}_x\uparrow}c_{j,\downarrow} + c^\dagger_{j-\hat{e}_x\uparrow}c_{j,\downarrow} + c^\dagger_{j+\hat{e}_y\uparrow}c_{j,\downarrow} + c^\dagger_{j-\hat{e}_y\uparrow}c_{j,\downarrow})$$
$$+ \mathcal{M}_3 c^\dagger_{j\uparrow}c_{j,\downarrow} \Big] + H.c. \tag{89}$$

$H_2$ is transformed to

$$H_2 = -\sum_j \sum_{\nu\nu'} J[\sigma_z]_{\nu\nu'}(-c^\dagger_{j+\hat{e}_x\nu}c_{j,\nu} + c^\dagger_{j+\hat{e}_y\nu}c_{j,\nu} + H.c.). \tag{90}$$

$H_2$ remains unchanged under Eq. (88). We make the following denotations

$$2\mathcal{M}_1 \equiv A_1, \qquad J \equiv A_2, \qquad \mathcal{M}_2 \equiv t_\parallel, \tag{91}$$

and assign the strength $\mathcal{M}_3$ with a tunable cyclical parameter $\theta$,

$$\mathcal{M}_3 = M_0 - 2t_\theta \cos(\theta). \tag{92}$$

Then Hamiltonian (73) gives

$$\begin{aligned} H(\mathbf{k}) = &\, 2A_2[\cos(k_x d) - \cos(k_y d)]\sigma_3 + 2A_1 \sin(k_x d)\sin(k_y d)\sigma_2 \\ &+ [M_0 - 2t_\| \cos(k_x d) - 2t_\| \cos(k_y d) - 2t_\theta \cos(\theta)]\sigma_1 + i\lambda\sigma_3. \end{aligned} \tag{93}$$

which is illustrated in Fig. 5. After recognizing $\theta$ as the artificial dimension $k_z$ and making the transformation $\sigma_3 \leftrightarrow \sigma_1$, one can find that Eq. (93) is identical to the model Hamiltonian (16) in the main text.

Finally, we discuss the relevant experiments related to the proposed cold-atom realizations. When the model Hamiltonians are engineered, existing techniques in cold atoms can provide various tools for exploring the focused physics of this work. For instance, the band structure of the model can be captured by the time-of-flight images [8], in which the nodal points are visualizable; the physics of the exceptional points introduced by the non-Hermitian terms can be extracted by the Rabi spectroscopy [7], and quantum behaviors such as Bloch oscillations [9] are readily investigated subject to the non-Hermiticity.



\end{document}